\documentclass[aps,prd,preprint,superscriptaddress]{revtex4-2}
\usepackage{graphicx} 
\usepackage{amsmath}
\usepackage{physics}
\usepackage{amssymb}
\usepackage{hyperref}
\usepackage{verbatim}
\usepackage{xcolor}

\begin{document}
\title{Holographic timelike entanglement and subregion complexity with scalar hair}
\date{\today}
\author{Hadyan Luthfan Prihadi}
\email{hadyanluthfanp9@gmail.com}
\affiliation{Research Center for Quantum Physics, National Research and Innovation Agency (BRIN),\\ South Tangerang 15314, Indonesia.}
\author{Muhammad Alifaldi Ramadhan Al-Faritsi}
\email{arexalif1@gmail.com}
\affiliation{Department of Engineering Physics, Institut Teknologi Bandung, Jl. Ganesha 10 Bandung, 40132, Indonesia.}
\author{Rafi Rizqy Firdaus}
\email{firdaus.rafi.rizqy.n7@s.mail.nagoya-u.ac.jp}
\affiliation{Department of Physics, Nagoya University,\\
Furo-cho, Chikusa-ku, Nagoya 464-8602, Japan}
\author{Fitria Khairunnisa}
\email{30223301@mahasiswa.itb.ac.id}
\affiliation{Theoretical High Energy Physics Group, Department of Physics, FMIPA, Institut Teknologi Bandung, Jl. Ganesha 10 Bandung, Indonesia.}
\author{Yanoar Pribadi Sarwono}
\email{yano001@brin.go.id}
\affiliation{Research Center for Quantum Physics, National Research and Innovation Agency (BRIN),\\ South Tangerang 15314, Indonesia.}
\author{Freddy Permana Zen}
\email{fpzen@fi.itb.ac.id}
\affiliation{Theoretical High Energy Physics Group, Department of Physics, FMIPA, Institut Teknologi Bandung, Jl. Ganesha 10 Bandung, Indonesia.}
\affiliation{Indonesia Center for Theoretical and Mathematical Physics (ICTMP), Institut Teknologi Bandung, Jl. Ganesha 10 Bandung,
	40132, Indonesia.}
\begin{abstract}
We investigate the holographic timelike entanglement entropy (HTEE) and timelike subregion complexity of a thermal CFT$_d$ deformed by a relevant scalar operator $\phi_0$, dual to a hairy black hole in AdS$_{d+1}$. We employ the prescription of merging spacelike and timelike surfaces at the interior, constructing an extremal surface homologous to a boundary timelike subsystem with a time interval $\Delta t$. Consequently, this deformation breaks the invariance of the imaginary component of HTEE observed in pure AdS$_3$ and BTZ geometry, introducing a nontrivial dependence on $\Delta t$. At small 
$\Delta t$, we derive analytical expressions that are in agreement with numerical results, and observe partial consistency with analytic continuation to temporal or spacelike entanglement entropy at the level of the near-boundary expansion. However, analytic continuation of CFT temporal entanglement entropy fails to reproduce the HTEE calculations under boundary deformation, even in $d=2$. Furthermore, we extend the numerical calculations to higher dimensions ($d=3$). In addition, we study holographic timelike subregion complexity within the complexity=volume conjecture and find that it remains real-valued, providing a complementary geometric probe of the black hole interior. In particular, for the BTZ black hole, we analytically show that the UV-finite term of the subregion complexity receives its entire contribution from the interior region alone.
\end{abstract}
\maketitle
\section{Introduction}
 Understanding the structure of spacetime behind black hole horizons remains one of the central challenges in gravitational physics. Recent developments show that various Kasner-like geometry near the black hole singularity emerge in a wide range of black hole solutions with a nontrivial scalar (or vector) hair \cite{Frenkel_2020,Cai2021,Arean2024,caceres2024kasnereon,cai2024,gao2024,Carballo2025,Zhang2025,cai2025clarifyingkasnerdynamicsinside,Prihadi2025b}. In the context of holographic theories realized through the AdS/CFT correspondence \cite{Maldacena1999}, the dual CFT theory is expected to encode not only the geometry outside the event horizon but also information about the black hole interior (see, for example, \cite{Heemskerk2012BulkTranshorizon,Hartman2013,Guo2025Duality}). Therefore, understanding how such interior structures are captured by boundary observables provides a valuable insight into the holographic description of spacetime beyond horizons. A concrete way to access this information is through extremal surfaces correspond to boundary observables, such as the holographic entanglement entropy \cite{Ryu2006PRL,Ryu2006,Hubeny2007}, which offer a geometric probe of the bulk spacetime, including the interior region \cite{Hartman2013}. Entanglement entropy is one of the most widely used measures of quantum correlations through entanglement, playing a central role across quantum information theory \cite{Horodecki2009} and many-body physics \cite{Islam2015}.\\
 \indent Recent studies have shown that timelike entanglement entropy \cite{Doi2023,Doi2023PRL,Afrasiar2024,Afrasiar2025, Heller2025,Heller2025PRX} and timelike subregion complexity \cite{Alishahiha2025} can be holographically calculated using generalized extremal surfaces anchored to the boundary timelike subregion $\mathcal{T}$, providing new tools to explore the interior region behind the horizon. Recently, study on the timelike entanglement entropy has attracted significant attention, which includes calculations in systems with Lorentz invariance breaking \cite{Afrasiar2024, Afrasiar2025,JenaMahapatra2025}, black hole background \cite{Heller2025,Heller2025PRX,afrasiar2025BH}, under $T\bar{T}$ deformation \cite{Jiang2023,Basu2024}, timelike entanglement first-law \cite{fujiki2025,li2025firstlaw}, applications to traversable wormhole \cite{Kawamoto2025Traversable,harper2025non}, and many more \cite{Li2023HTEE, JiangWangWuYang2023,ChuParihar2023,Grieninger2024,NarayanSaini2024,HeZhang2024,Liu2024Entang,Nunez2025,XuGuo2025,Guo2025Relation,ChuParihar2025,giataganas2025}. While timelike entanglement entropy has been extensively studied in vacuum AdS and simple black hole backgrounds, much less is known once matter fields deform the geometry and generate a nontrivial Kasner interior.\\
 \indent In $d=2$, the holographic calculations proposed in \cite{Afrasiar2024} and \cite{Heller2025} match the analytic continuation of the CFT entanglement entropy to timelike subregion. Furthermore, the imaginary part of the timelike entanglement entropy is a constant in $d=2$. However, it remains unclear whether the timelike extremal surfaces continue to provide a faithful gravitational dual of the analytically continued CFT entanglement entropy in the presence of relevant boundary deformations. In particular, relevant deformations are known to break conformal symmetry and modify the IR structure of the theory \cite{Frenkel_2020,Caceres2022TransIR}.  We aim to investigate how the holographic calculations of timelike entanglement entropy and timelike complexity are influenced by boundary relevant deformations. Note that our primary objective is not merely to extend existing computations to a deformed background, but to address a more important question: to what extent do the extremal surfaces anchored to $\mathcal{T}$ remain reliable as geometric representatives of the analytically continued CFT entanglement entropy, and how they behave as probes of black hole interiors once the bulk geometry is nontrivially deformed?\\
\begin{figure}
    \centering
    \includegraphics[width=0.85
\linewidth]{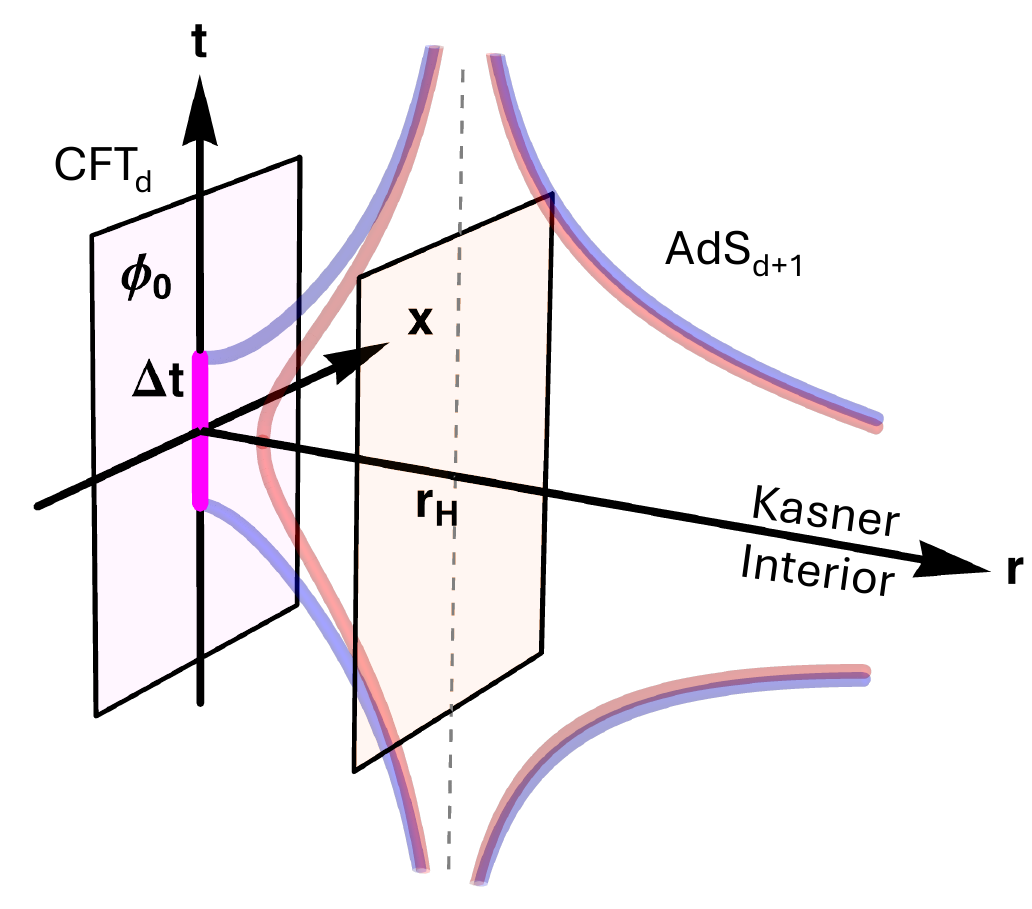}
    \caption{Illustration of the AdS$_{d+1}$ black brane setup where the CFT$_d$ is deformed by $\phi_0$ (higher dimensions are suppressed). The timelike subregion $\mathcal{T}$ is expressed as a strip with time interval $\Delta t$ localized at $x=0$. Here, $r$ is the holographic direction, with the asymptotic boundary at $r=0$ and the horizon at $r=r_H$. The blue curves denote spacelike extremal surfaces (real area), while the red curves denote timelike ones (imaginary area). The two surfaces merge at $r\rightarrow\infty$.}
    \label{fig:setup}
\end{figure}
\indent In this work, we calculate the timelike entanglement entropy of a thermal CFT$_d$ deformed by a scalar source $\phi_0$, using the holographic prescription. On the gravity side, $\phi_0$ specifies the boundary condition of a bulk Klein-Gordon scalar field $\phi(r)$ at the asymptotic boundary $r\rightarrow0$. The scalar field propagates in a AdS$_{d+1}$ black hole/black brane background and induces a significant backreaction on the geometry. While the deformation may be small near the boundary, it has a strong backreaction in the bulk as the deformation is relevant. In particular, it substantially modifies the black hole interior into a more general Kasner universe, whose Kasner exponents are determined by the value of $\phi_0$. Several previous works have calculated the holographic entanglement entropy for this hairy black hole background using the Ryu-Takayanagi formula \cite{Frenkel_2020, Caceres2024}, including in the holographic superconductors \cite{Hartnoll2008,Hartnoll2021} where the scalar field is coupled with a Maxwell gauge field (see, for example, \cite{Cai2012,Cai2012b, Albash_2012, GarciaGarcia2015,li2015entanglemententropyholographicpwave,Prihadi2025a, Prihadi2025b}). Our gravity setup for the holographic timelike entanglement entropy calculation is illustrated in figure \ref{fig:setup}.\\ 
\indent The timelike entanglement entropy can be computed holographically in a variety of methods such as calculating extremal surfaces in a complexified manifold \cite{Heller2025,Heller2025PRX}. Here, we calculate the holographic timelike entanglement entropy (HTEE) using the surface-merging prescription, namely we merge the spacelike and timelike extremal surfaces as previously done in \cite{Afrasiar2024,Afrasiar2025,giataganas2025}, and recently in \cite{afrasiar2025BH} for black holes in AdS. In this case, the spacetime coordinates and the area of the spacelike surface remain real-valued, while the area of the timelike surface is purely imaginary. Furthermore, we also keep the Lorentzian signature of the spacetime metric throughout the calculations. This method also reduces to the known results in \cite{Doi2023} for vacuum AdS and for a $d=2$ Ba$\tilde{\text{n}}$ados-Teitelboim-Zanelli (BTZ) black hole, although the result for $d>2$ black holes and their relation with the findings in \cite{Heller2025,Heller2025PRX} remains elusive. \\ 
\indent Interestingly, both spacelike and timelike surfaces penetrate the horizon and eventually merge as $r\rightarrow\infty$. This near-singularity merging behavior motivates our investigation into how the timelike entanglement entropy of the boundary theory captures the Kasner geometry generated by the relevant deformation $\phi_0$. Therefore, it is important to study the behavior of HTEE under such a deformation. As $\phi_0$ modifies both the exterior and interior structure, we anticipate that the resulting HTEE will exhibit nontrivial behavior under this deformation. Furthermore, explicit calculations in AdS$_3$ could serve as a test of whether this holographic prescription continues to agree with results obtained via analytic continuation from spacelike or temporal entanglement.\\
\indent In addition, we also calculate the holographic timelike subregion complexity recently proposed in \cite{Alishahiha2025}, as a timelike extension to the previous holographic subregion complexity proposal \cite{Alishahiha2015,BenAmi2016,Carmi2017}. The conjecture states that the complexity of a timelike subregion $\mathcal{T}$ is proportional to the volume enclosed by the extremal surface $\gamma_\mathcal{T}$, within the complexity = volume (CV) conjecture \cite{Stanford2014CV,Susskind2016CV}. This provides a holographic calculation for a measure of complexity associated with the time evolution in the boundary theory. Unlike HTEE, the holographic timelike complexity remains purely real, allowing for an unambiguous geometric interpretation. As the extremal surface penetrates the black hole horizon, the associated volume extends deep into the Kasner interior as well. Therefore, the holographic timelike complexity may give us a clear geometric picture of the black hole interior from a quantum information perspective, one that is likely inaccessible to the other holographic entanglement measures. We study how this volume is influenced by the boundary deformation $\phi_0$ and see how black hole interior plays a role in determining the subregion complexity.\\
\indent We outline the structure of the paper and summarize our main results. In Section \ref{sec2}, we present the gravitational setup of an AdS$_{d+1}$ black hole background with a scalar field $\phi(r)$, which was previously studied in \cite{Caceres2024}. In Section \ref{sec3}, we calculate the HTEE in this background by merging spacelike and timelike extremal surfaces. We demonstrate that such a merger is possible in the Kasner interior, even in the presence of a scalar backreaction. We present our numerical calculations to the HTEE and conclude that the scalar field generically enhances the real part while suppressing the imaginary part of the HTEE. We also find that the imaginary part of HTEE depends on $\Delta t$ once the scalar hair is turned on, even in AdS$_3$. We obtain analytical expressions for near-vacuum case at small $\Delta t$, which agrees with the numerical calculations under data fitting. While reverse analytic continuation from HTEE to spacelike or temporal entanglement entropy reproduces part of the near-boundary behavior, we argue that discrepancies arise from deeper IR contributions. Importantly, we find that the analytic continuation of CFT entanglement entropy fails to reproduce the deformed HTEE calculations, even in $d=2$. We also provide a brief comment on black holes with inner horizons. In Section \ref{sec4}, we calculate the holographic timelike subregion complexity and find that the subleading term grows linearly in $\Delta t$ before eventually saturating. We also analytically show that, in the BTZ background, the linear growth is governed by the volume inside the horizon. In Section \ref{sec5}, we present discussions of the results.
\section{AdS Black Holes with Scalar Hair}\label{sec2}
In this section, we briefly review the AdS$_{d+1}$ black hole solution with a massive real scalar hair $\phi(r)$ with mass $m$ studied in \cite{Caceres2024} in the context of holographic renormalization group (RG) flow. The action of this theory in a $(d+1)$-dimensional Einstein gravity is given by
\begin{equation}
    S=\frac{1}{\kappa^2}\int d^{d+1}x\sqrt{-g}\bigg(R-2\Lambda-\frac{1}{2}(\nabla\phi)^2-\frac{1}{2}m^2\phi^2\bigg),
\end{equation}
where $\kappa^2=16\pi G_N^{d+1}$ is the gravitational coupling and $\Lambda=-\frac{d(d-1)}{2L^2}$ is the negative cosmological constant for AdS spacetime.\\
\indent The metric ansatz for this hairy black hole is given by
\begin{equation}\label{metrichairy}
    ds^2=\frac{L^2}{r^2}\bigg(-fe^{-\chi}dt^2+\frac{dr^2}{f}+dx^2+d\textbf{y}_{d-2}^2\bigg),
\end{equation}
where $L$ is the AdS radius and $d$ represents the CFT$_d$ spacetime dimension. Here, we impose the AdS boundary condition in the asymptotic boundary located at $r=0$, where $f(0)=1$ and $\chi(0)=0$. The black hole's horizon is located at $r=r_H$ defined from $f(r_H)=0$, where the singularity is located at $r\rightarrow\infty$. For a Scwarzschild-AdS$_{d+1}$ spacetime, we have $f(r)=1-(r/r_H)^{d}$ and $\chi(r)=\phi(r)=0$. In the numerical calculations, we can rescale the fields to set $L=1$ and $r_H=1$ without altering the form of the equations of motion.\\
\indent The role of the scalar field is to deform the CFT theory as
\begin{equation}
    \delta S_{\text{CFT}}\sim\int d^dx \phi_0\langle\mathcal{O}\rangle,
\end{equation}
where $\phi_0$ is the source of the deformation and the expectation value of the single-trace operator $\mathcal{O}$ is the corresponding response with a conformal dimension $\Delta$. This deformation generates a RG flow from the UV boundary toward the IR region near the horizon, and possibly extend into a trans-IR regime in the black hole interior \cite{Caceres2022TransIR}. Here, we consider a relevant deformation with $-\frac{d^2}{4}<m^2<0$ ($\Delta<d$) which also satisfy the Breitenlohner-Friedmann bound.\\
\indent The metric functions $f(r)$ and $\chi(r)$ along with the scalar field $\phi(r)$ solve the equations of motion coming from the Klein-Gordon equation and the $tt$- and $rr$-components of the Einstein's equation. We solve the equations of motion numerically subject to the AdS boundary condition, where the scalar field behaves as $\phi(r)\sim\phi_-r^{d-\Delta}+\phi_+r^{\Delta}$ near the boundary with $\Delta=\frac{d}{2}+\sqrt{\frac{d^2}{4}+m^2}$. In the context of the AdS/CFT correspondence \cite{deHaro2001,Skenderis2002}, the identification of the source $\phi_0$ to the boundary deformation depends on the value of the scaling dimension $\Delta$, hence the value of $m^2$. Depending on whether we choose the standard or alternative quantization, we have
\begin{equation}\label{boundaryconditions}
    \phi_0=\begin{cases}
        \lim_{r\rightarrow 0}r^{\Delta -d}\phi(r),\;\;\;&\Delta >\frac{d}{2},\\
        \lim_{r\rightarrow 0}-\frac{r^{2\Delta -d+1}}{2\Delta -d}\partial_r(r^{-\Delta }\phi(r)),\;\;\;&\Delta <\frac{d}{2},\\
        \lim_{r\rightarrow 0}-\frac{r^{-d/2}}{\log r}\phi(r),\;\;\;&\Delta =\frac{d}{2}.
    \end{cases}
\end{equation}
The last boundary condition facilitates the appearance of a logarithmic factor in the scalar field near the boundary when $\Delta=\frac{d}{2}$.\\
\indent To find the numerical solutions of the fields $\{\phi,f,\chi\}$, we integrate the equations of motion numerically from $r=r_H-\delta$ to $r=\varepsilon$ and from $r=r_H+\delta$ to $r\rightarrow\infty$, where $\delta=10^{-6}$ and $\varepsilon=10^{-10}$ are chosen throughout the numerical calculations. The fields $\{\phi,f,\chi\}$ are expanded near $r=r_H$ and the expansion coefficients are chosen so that the equations of motion do not diverge at the horizon. All of the near-horizon expansion coefficients can be determined by the value of $f^\prime(r_H)$, which defines the temperature of the black hole,
\begin{equation}
    T=\frac{|f^\prime(r_H)|e^{-\chi(r_H)/2}}{4\pi}.
\end{equation}
\indent When there is no scalar field, $f^\prime(r_H)=-\frac{d}{r_H}$. Therefore, by gradually making $f^\prime(r_H)$ more negative, we simultaneously increase the value of the boundary scalar field $\phi_0$, which is obtained from Eq. \eqref{boundaryconditions}, depending on the boundary conditions of $\phi(r)$. In this work, we express the boundary deformation parameter as its dimensionless form $\tilde{\phi}_0\equiv T^{-d+\Delta}\phi_0$.
\section{Holographic Timelike Entanglement Entropy}\label{sec3}
\begin{figure}
    \centering    \includegraphics[width=0.47\linewidth]{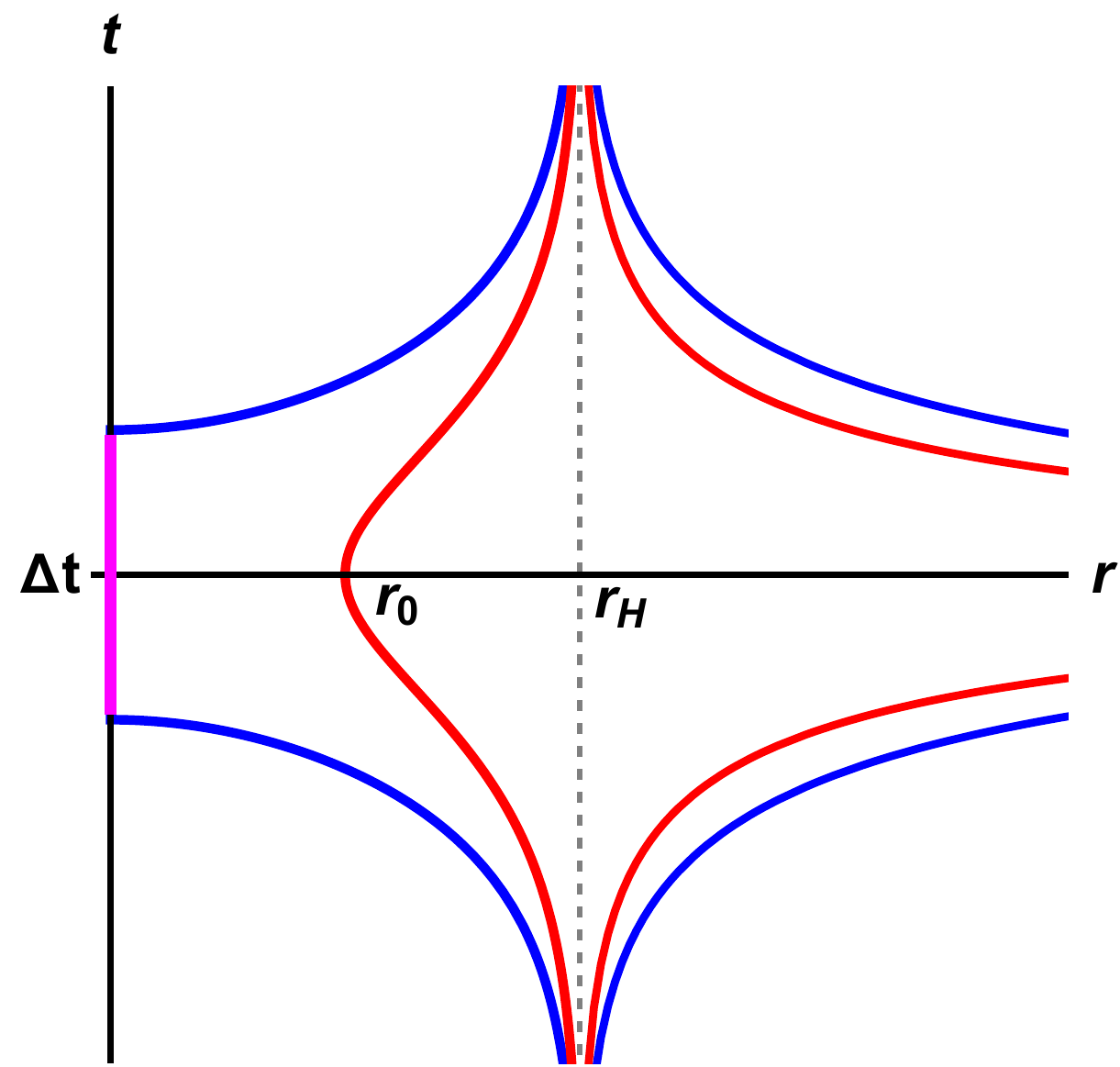}
    \includegraphics[width=0.47\linewidth]{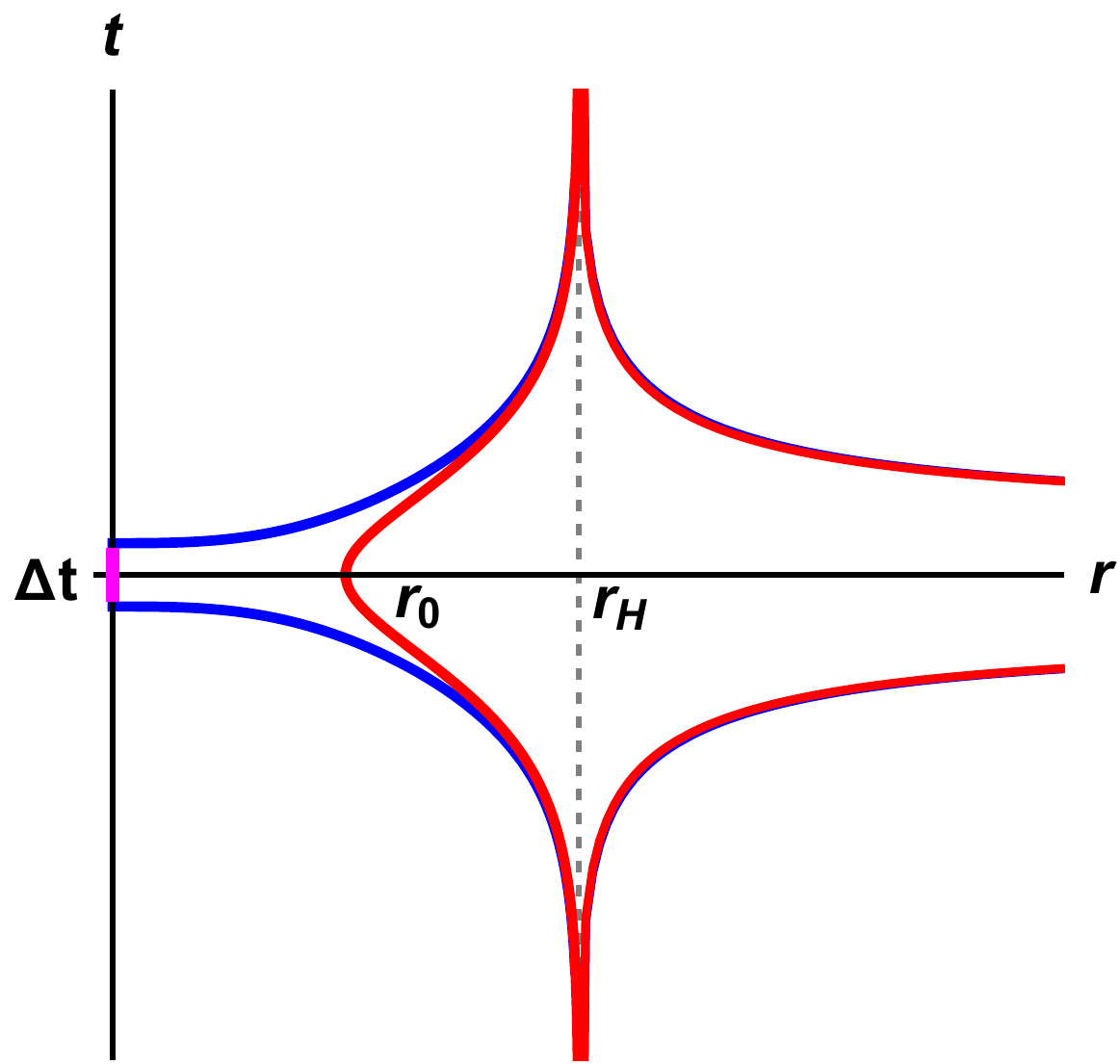}
    \caption{Illustration of the $s=+1$ surfaces (blue) and $s=-1$ surfaces (red) for $\tilde{\phi}_0=0$ (left) and $\tilde{\phi}_0=2.1533$ (right). Here we choose $d = 2$, $\Delta =1.5$, $r_H=1$, and $r_0=0.5$.}
    \label{fig:HTEE}
\end{figure}
\indent The HTEE associated with a timelike subregion $\mathcal{T}$ of the CFT$_d$ is proposed to be given by \cite{Doi2023PRL,Doi2023,Afrasiar2024,Afrasiar2025,Heller2025,Heller2025PRX}
\begin{equation}
S_{\mathcal{T}}=\frac{\mathcal{A}(\gamma_\mathcal{T})}{4G_N^{(d+1)}},
\end{equation}
where $\mathcal{A}(\gamma_\mathcal{T})$ denotes the area of a co-dimension two extremal surface $\gamma_\mathcal{T}$ in AdS$_{d+1}$ anchored to the boundary of $\mathcal{T}$. In this case, $\gamma_\mathcal{T}$ is the combination of spacelike and timelike surfaces that are merged together close to the black hole singularity. In this work, we choose $\mathcal{T}$ as a simple timelike strip with interval $\Delta t$,
\small
\begin{equation}
    \mathcal{T}=\bigg\{(t,x,\textbf{y}_{d-2}):t\in\bigg[-\frac{\Delta t}{2},\frac{\Delta t}{2}\bigg],x=0,\textbf{y}_{d-2}\in\mathbb{R}^{d-2}\bigg\}.
\end{equation}
\normalsize
\indent In the hairy black hole background where the metric is given by Eq. \eqref{metrichairy}, the area functional is given by
\begin{equation}\label{areafunctional}
    \mathcal{A}(\gamma_\mathcal{T})=V_{d-2}L^{d-1}\int\frac{dr}{r^{d-1}}\sqrt{-fe^{-\chi}t^{\prime 2}+\frac{1}{f}},
\end{equation}
where $V_{d-2}$ denotes the spatial volume of the $\textbf{y}_{d-2}$ coordinates and $t^\prime\equiv\frac{dt(r)}{dr}$. After minimizing the area functional, we obtain
\begin{equation}\label{tprimesquare}
    t^{\prime 2}(r)=\frac{K^2 r^{2d-2}}{f^2e^{-\chi}}\frac{1}{fe^{-\chi}+K^2r^{2d-2}},
\end{equation}
where $K^2$ is the associated conserved quantity. The important difference between $\mathcal{A}(\gamma_\mathcal{T})$ for a timelike subregion $\mathcal{T}$ and $\mathcal{A}(\gamma_A)$ for a spacelike subregion $A$ in a hairy black hole background is that $\mathcal{A}(\gamma_A)$ does not explicitly depends on $\chi(r)$, which makes HTEE more sensitive to the presence of the scalar hair.\\
\indent There are several possible cases that can provide the extremal surface used to compute the timelike entanglement entropy. The first one is when $K^2>0$, which leads to spacelike extremal surfaces with real-valued area. The second one is $K^2<0$, yielding timelike extremal surfaces whose areas become purely imaginary. Both spacelike and timelike surfaces are then merged to form a continuous extremal surface homologous to the boundary time interval. A surface with $K^2=0$ ($t(r)=\frac{\Delta t}{2}$ stretches from $r=0$ to $r\rightarrow\infty$) can also be considered to regulate the spacelike surface which has a UV divergence. Thus, one can define $K^2=s\tilde{K}^2$, where $s=\pm1$ \cite{Afrasiar2025}, and calculate the total area of all spacelike and timelike extremal surfaces.\\
\indent From dimensional analysis, we can write the constant $\tilde{K}^2$ as
\begin{equation}\label{Ktilde}
    \tilde{K}^2=\frac{f_0e^{-\chi_0}}{r_0^{2d-2}},
\end{equation}
where $f_0\equiv f(r_0)$ and $\chi_0\equiv \chi(r_0)$ and $r_0<r_H$. As long as the turning point remains outside the horizon, the value of $\tilde{K}^2$ remains positive. From this, we can write $t^\prime$ that extremizes the area functional as
\begin{equation}\label{tprimes}
    t^\prime_{s}(r)=\frac{e^{\chi/2}}{f}\bigg(1+s\bigg(\frac{r_0}{r}\bigg)^{2d-2}\frac{fe^{-\chi}}{f_0e^{-\chi_0}}\bigg)^{-\frac{1}{2}},
\end{equation}
with $s=+1$ ($t_+(r)$) for the spacelike surfaces and $s=-1$ ($t_-(r)$) for the timelike surfaces.\\
\indent Note that one can define $r_0$ as the turning point for the timelike surface $t_-(r)$ outside the horizon such that $t^\prime_{-}(r_0)\rightarrow\infty$. The $t_-(r)$ surface starts from $t_{-}(r_0)=0$ at the boundary to $t_{-}(r_H)\rightarrow\infty$ at the horizon. To extend this solution into the black hole interior, we identify the exterior and interior segments by imposing $t_-(r_H-\delta)=t_-(r_H+\delta)$. Starting from this matched value at $r = r_H + \delta$, we integrate the interior solution of $ t_-(r)$ toward larger $ r$, up to $ r\rightarrow \infty$, which corresponds to the location of the singularity. In the numerical calculations, we put the singularity at a large but finite $r = 1/\delta$.\\
\indent This timelike surface is then merged with the spacelike surfaces in the vicinity of the singularity. The spacelike surface $t_+(r)$ starts from $r\rightarrow\infty$ to $r=r_H+\delta$. We merge the spacelike and timelike surfaces by identifying $t_+(\infty)=t_-(\infty)$. Furthermore, at the horizon we smoothly connect the two sides of the spacelike surface by matching $t_+(r_H+\delta)=t_+(r_H-\delta)$, ensuring a continuous function across the horizon. The exterior solution for $t_+(r)$ is obtained from $r=r_H-\delta$ down to the asymptotic boundary at $r=\varepsilon$. The boundary time interval is then recovered from the relation $\Delta t=2t_+(\varepsilon)$.\\
\indent The spacelike and timelike extremal surfaces for AdS$_3$ is depicted in figure \ref{fig:HTEE}. Once the spacelike and timelike surfaces are joined, they form a single continuous surface homologous to the boundary time interval $\Delta t$. This surface can then be treated as an extremal surface that is relevant to compute the timelike entanglement entropy holographically \cite{afrasiar2025BH}. By initially fixing the value of $r_0>r_H$ and merging the surfaces as explained before, we allow the boundary time interval $\Delta t$ to be controlled by the value of $r_0$. For a three-dimensional BTZ black hole with $f(r)=1-\big(\frac{r}{r_H}\big)^2$ and $\chi(r)=0$, the HTEE matches the results obtained by analytic continuation from a spacelike interval $\Delta x$ to $\Delta x\rightarrow i\Delta t$ \cite{Doi2023,Doi2023PRL,Heller2025}, which is given by (see Appendix A for more detailed derivation)
\begin{equation}\label{analyticalarea}
    \mathcal{A}(\gamma_\mathcal{T})\big|_{\phi_0=0}=2L\ln\bigg(\frac{2r_H}{\varepsilon}\sinh\bigg(\frac{\Delta t}{2r_H}\bigg)\bigg)+i\pi L,
\end{equation}
where the relation between $r_0$ and $\Delta t$ is given by
\begin{equation}\label{dtanalytical}
    \Delta t\big|_{\phi_0=0}=2r_H\tanh^{-1}\bigg(\frac{(r_0/r_H)}{\sqrt{1-(r_0/r_H)^2}}\bigg).
\end{equation}
\subsection{Extremal Surfaces Merging Near Kasner Singularity}
Another important aspect of the hairy black hole solution is the emergence of a more general Kasner spacetime near the singularity. This subsection is included as a consistency check, verifying that the spacelike–timelike extremal surface construction remains smooth and well defined in this Kasner regime. Although the scalar field induces only small deformations outside the horizon, it grows without a bound inside the black hole, leading to a substantial modification of the interior geometry. At $r\gg r_H$, the solutions to the equations of motion can be obtained analytically, which give us
\begin{align}\label{fieldskasner}
    \phi(r)\sim(d-1)c_\phi\log r+\phi_K,&\;\;\; f(r)\sim-f_Kr^{\big(d+\frac{d-1}{2}c_\phi^2\big)},\;\;\;\\\nonumber
    \chi(r)\sim(d-1)&c_\phi^2\log r+\chi_K,
\end{align}
where $c_\phi,\phi_K,f_K,\chi_K$ are the near-singularity integration constants. With this expression, the metric near the singularity becomes a Kasner spacetime
\begin{align}
    ds^2=&-d\tau^2+\tau^{2p_t}dt^2+\tau^{2p_x}(dx^2+d\textbf{y}_{d-2}^2),\\
    \phi(r)=&-\sqrt{2}p_\phi\log\tau,
\end{align}
after performing a coordinate transformation $\tau=r^{-\frac{1}{2}\big(d+\frac{d-1}{2}c^2_\phi\big)}$. Here, $p_t,p_x,p_\phi$ are the Kasner exponents defined as
\begin{align}
    p_t=&\frac{(d-1)c_\phi^2-2(d-2)}{2d+(d-1)c_\phi^2},\\
    p_x=&\frac{4}{2d+(d-1)c_\phi^2},\\
    p_\phi=&\frac{2\sqrt{2}(d-1)c_\phi}{2d+(d-1)c_\phi^2}.
\end{align}
\indent The sign of the Kasner exponent $p_t$ determine whether the spacetime grows or shrink as it approach singularity. When $p_t<0$ (equivalent to $c_\phi^2<\frac{2(d-2)}{(d-1)}$), the term $\tau^{2p_t}$ grows as we approach the singularity. However, the sign might be flipped to $p_t>0$ when the geometry undergoes a Kasner inversion, causing a change $c_\phi^2\rightarrow\frac{1}{c_\phi^2}$. This inversion (or transition) might occur, for example, due to the coupling between the scalar hair to a Maxwell field $A_\mu$, or even when the spacetime is stationary \cite{gao2024,Prihadi2025b}.\\
\indent In this regime, the merger between spacelike and timelike surfaces occur. Therefore, it is important to learn how $t^\prime_s(r)$ is influenced by the Kasner exponent. In \cite{Afrasiar2024,Afrasiar2025}, it is argued that a smooth transition between spacelike and timelike hypersurfaces requires the induced norms of the two extremal surfaces, $g^{\mu\nu}\partial_\mu\Sigma_\text{Re}\partial_\nu\Sigma_\text{Re}$ and $g^{\mu\nu}\partial_\mu\Sigma_\text{Im} \partial_\nu\Sigma_\text{Im}$ are equal, where $\Sigma_\text{Re}$ and $\Sigma_\text{Im}$ denote the spacelike and timelike hypersurfaces, respectively. In our setup, this requirement translates into demanding that the squared derivatives $(t_+^\prime)^2$ and $(t_-^{\prime})^2$ are equal in the Kasner singularity. Since $(t_+^\prime)^2>0$ and $(t_-^\prime)^2<0$ in the interior, both need to vanish as $r\rightarrow\infty$.\\
\indent We could explicitly check how the extremal surfaces gets affected by the Kasner exponent. Near the singularity where $r\rightarrow\Lambda\gg1$, using the expressions in Eq. \eqref{fieldskasner}, $t^\prime_s(r)$ can be written as
\small
\begin{equation}
    t_s^\prime (r)^2\sim\frac{e^{\chi_K}}{f_K^2\Lambda^{2d}}\bigg(1-s\bigg(\frac{r_0^{2d-2}}{f_0e^{-\chi_0}}\bigg)f_Ke^{-\chi_K}\Lambda ^{-[(d-2)+\frac{d-1}{2}c_\phi^2]}\bigg)^{-1}.
\end{equation}
\normalsize
Since $c_\phi^2>0$ and $d\geq2$, the exponent of $\Lambda$ in the last term is always negative. In $d=2$, the exponent is entirely determined by $c_\phi^2$. Therefore, as $\Lambda\rightarrow\infty$, $t^\prime_s(r)$ always vanishes for any $s=\pm1$. As a result, the spacelike and timelike segments of the extremal surface can be smoothly merged in the Kasner regime, no matter how strong the scalar backreaction is. Importantly, this statement remains valid even if the model undergoes a Kasner inversion/transition, as long as $c_\phi^2>0$. Allowing $c_\phi^2<0$ would correspond to a phantom-like scalar violating energy conditions, and is not considered here.
\begin{figure}
    \centering
\includegraphics[width=0.9\linewidth]{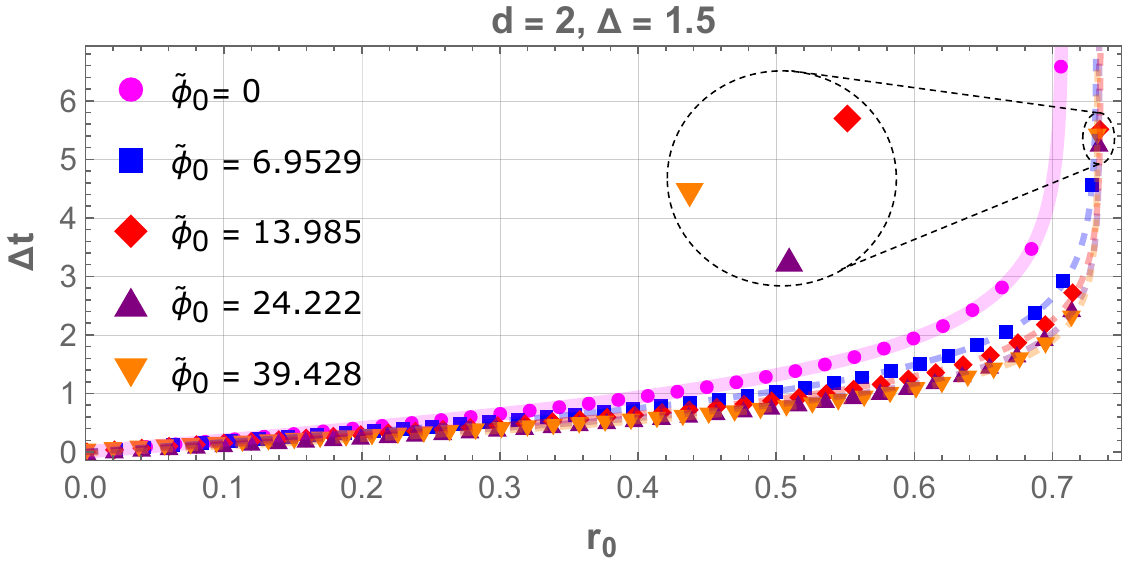}\\\includegraphics[width=0.9\linewidth]{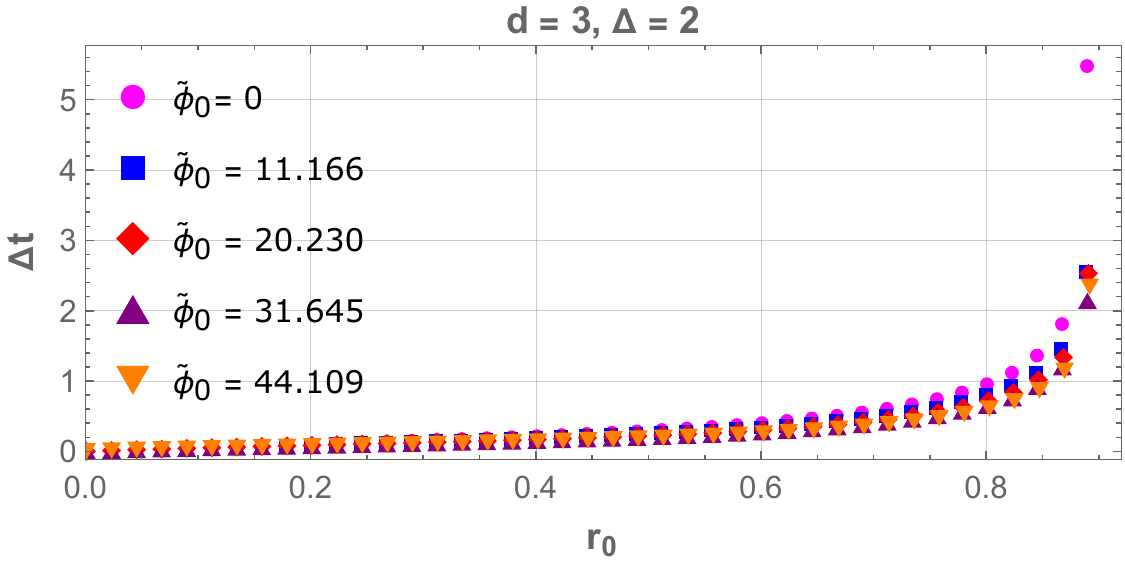}
    \caption{Relations between boundary time interval $\Delta t$ and the turning point $r_0$ for various dimensions and $\tilde{\phi}_0$. Magenta solid line represents the analytical solution for $d=2$, which is given by Eq. \eqref{dtanalytical}. The colored dashed lines are the fitting function in Eq. \eqref{fittinginterval}.}
    \label{fig:dt}
\end{figure}
\subsection{Numerical Results}
\begin{figure*}
    \centering
    \includegraphics[width=0.45\linewidth]{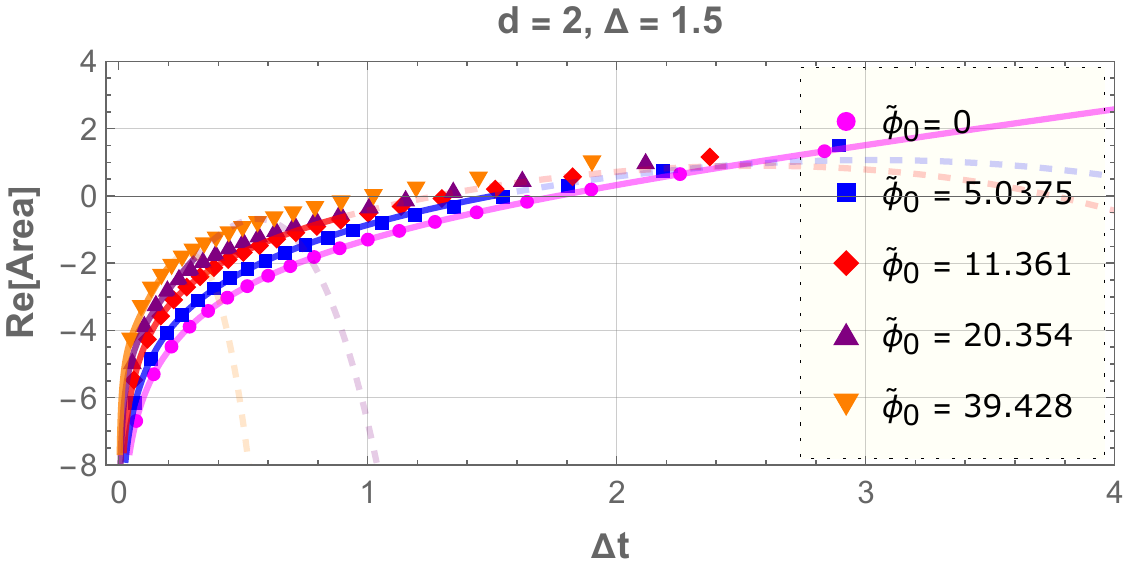}\;\;\;\;\;
    \includegraphics[width=0.45\linewidth]{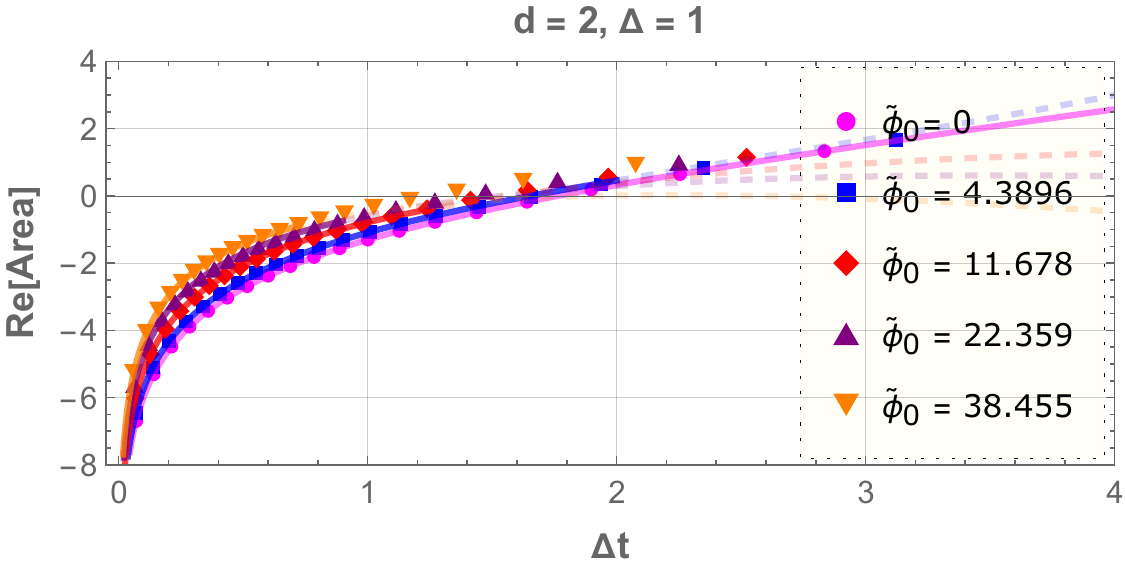}\\
    \includegraphics[width=0.45\linewidth]{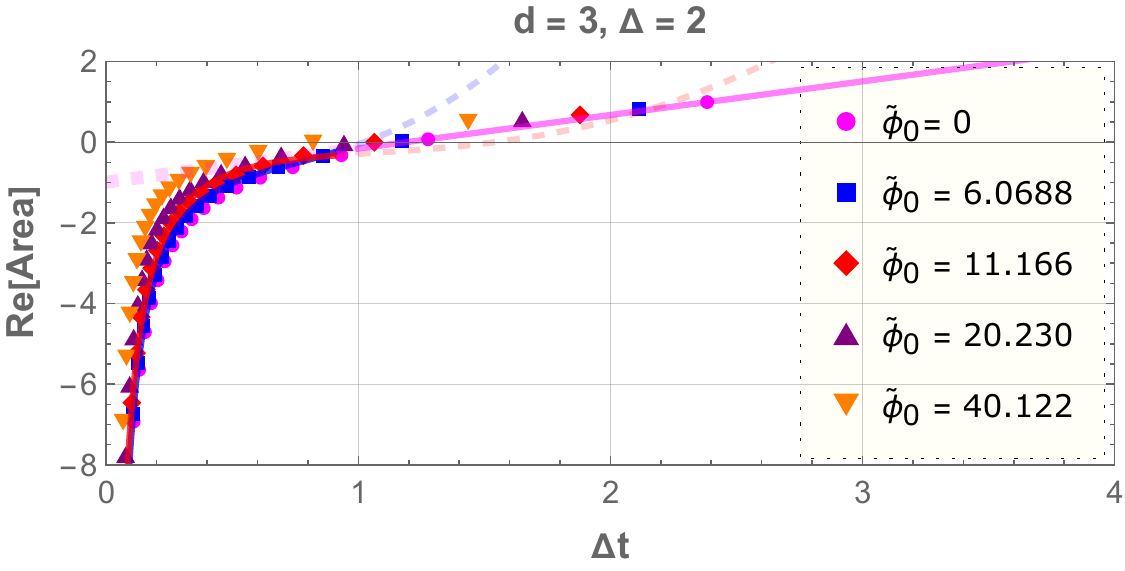}\;\;\;\;\;
    \includegraphics[width=0.45\linewidth]{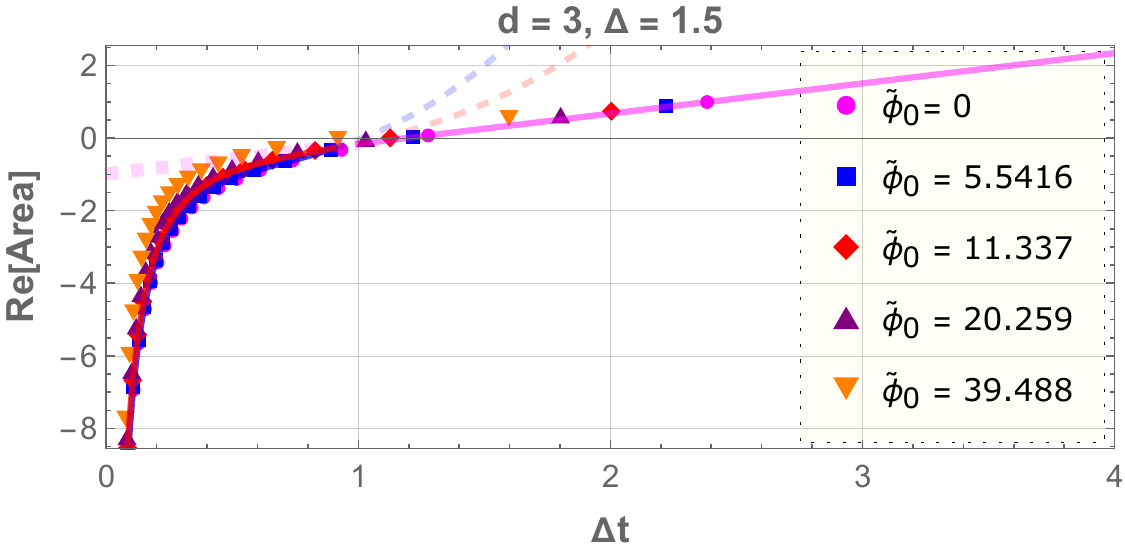}\\
    \caption{Real part of the total area versus the boundary time interval $\Delta t$ for various boundary deformation parameter $\tilde{\phi}_0=T^{-d+\Delta}\phi_0$. The solid magenta line represents analytical result from Eq. \eqref{analyticalarea} for $d=2$ and solid colored lines represent fitted functions from Eqs. \eqref{fitd2D32} and \eqref{fitd2D1}. In $d=3$ cases, the solid magenta line represents a linear function of $\Delta t$, fitted with the $\tilde{\phi}_0$ results while colored lines represent fitted functions from Eqs. \eqref{fitd3D2} and \eqref{fitd3D32}. However, our numerical fitting breaks down at large $\tilde{\phi}_0$ for $d=3$ due to the breakdown of the near-boundary approximation.}
    \label{fig:PlotRe}
\end{figure*}
\begin{figure*}
    \centering
    \includegraphics[width=0.45\linewidth]{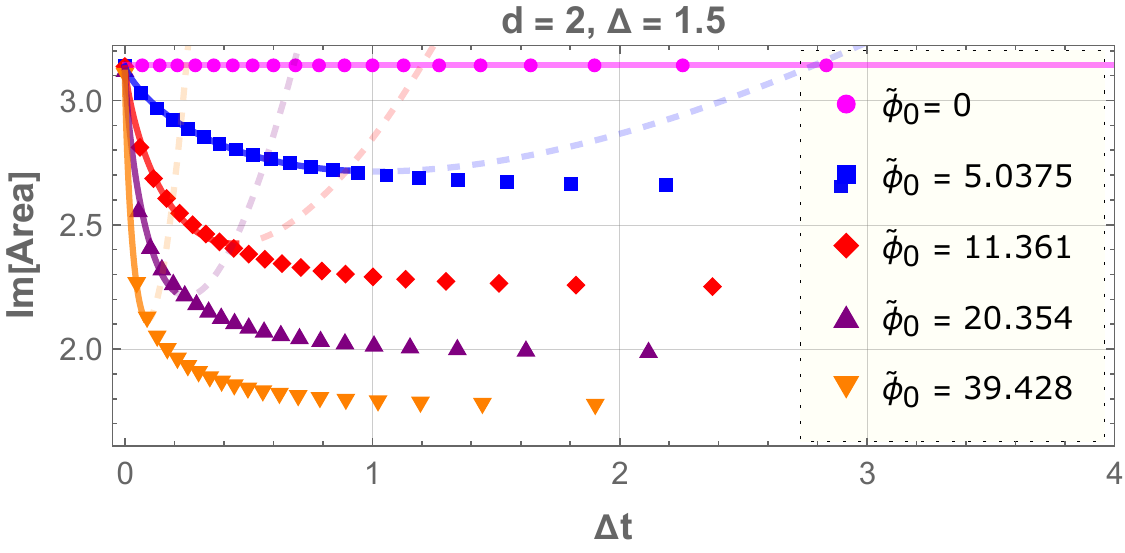}\;\;\;\;\;
    \includegraphics[width=0.45\linewidth]{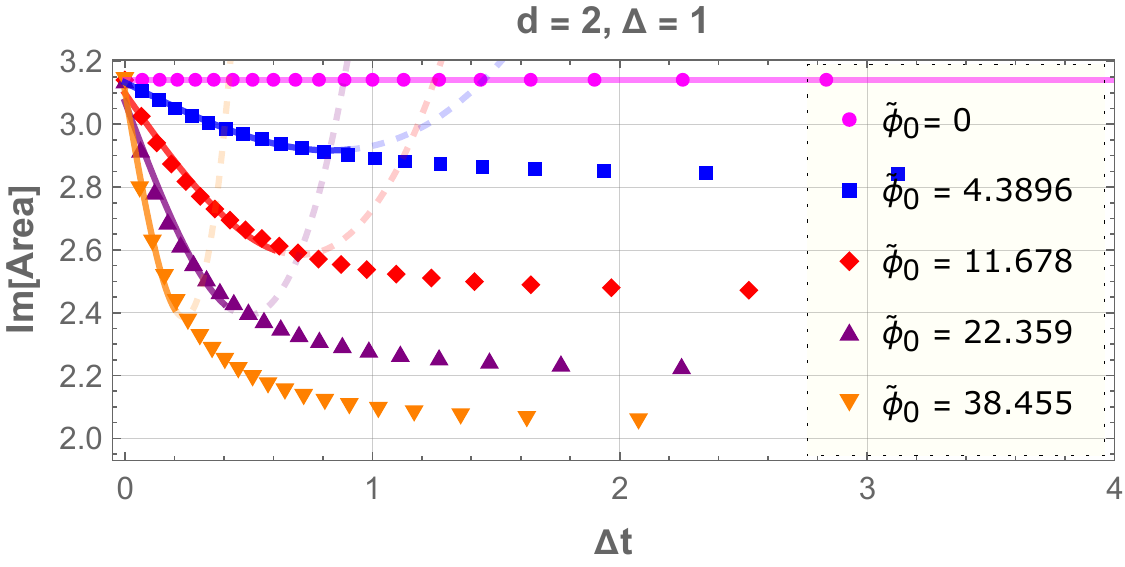}\\
    \includegraphics[width=0.45\linewidth]{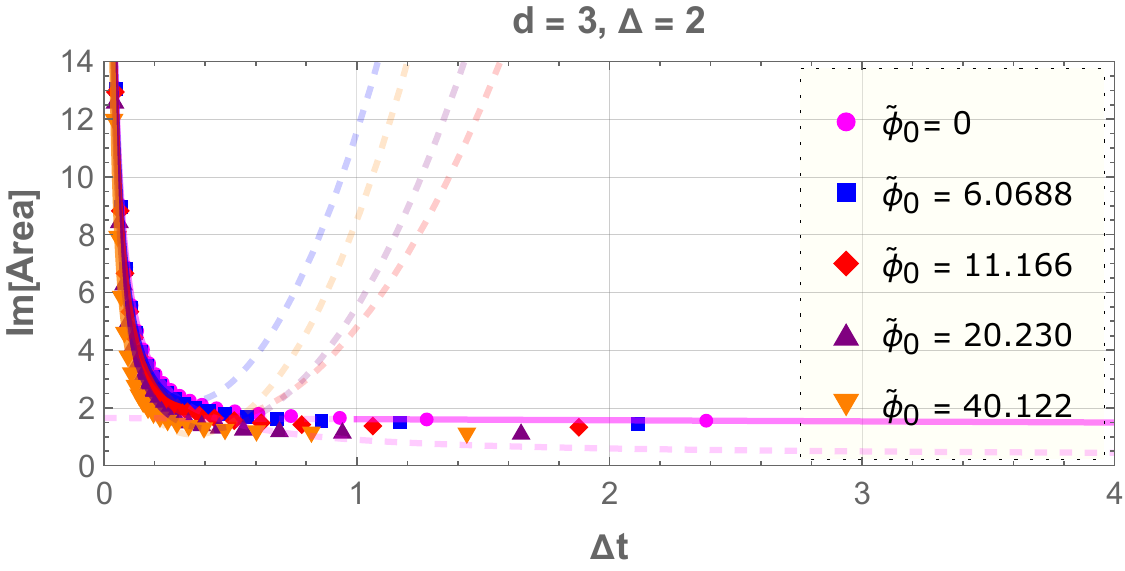}\;\;\;\;\;
    \includegraphics[width=0.45\linewidth]{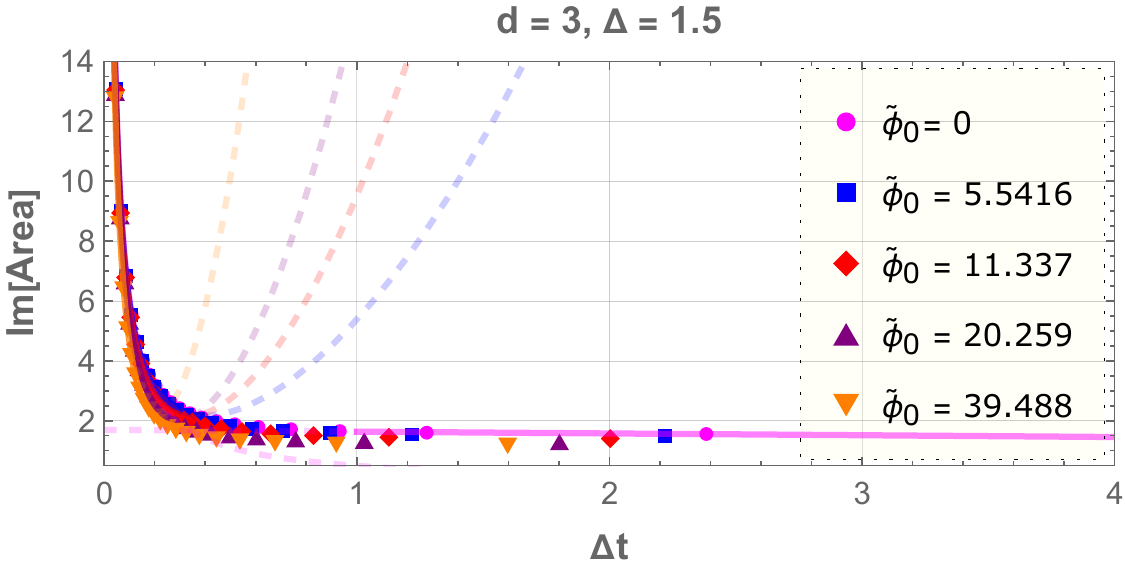}\\
    \caption{Imaginary part of the total area versus the boundary time interval $\Delta t$ for various boundary deformation parameter $\tilde{\phi}_0=T^{-d+\Delta}\phi_0$. For $d=2$, the solid magenta line represents analytical result from Eq. \eqref{analyticalarea} and solid colored lines represent analytical fitting from Eqs. \eqref{fitd2D32} and \eqref{fitd2D1} for small $\Delta t$. In $d=3$, we fit the functions presented in Eqs \eqref{ImaginaryFit1} and \eqref{ImaginaryFit2} to the $\tilde{\phi}_0=0$ results.}
    \label{fig:PlotIm}
\end{figure*}
\indent To calculate the area with scalar field, we use the numerical solutions to $f(r)$ and $\chi(r)$ from the equations of motion and numerically integrate the area functional. The spacelike surface outside the horizon is regularized by a UV cut-off $\varepsilon$ where we choose $\varepsilon=10^{-10}$ in the numerical calculations. Furthermore, we vary $r_0$, which leads to the variation of $\Delta t$. We then gradually increase the value of the boundary deformation $\phi_0$, which is represented as its dimensionless value $\tilde{\phi}_0\equiv T^{-d+\Delta}\phi_0$. For $d=2$, we consider cases with $\Delta = 1.5$ and $\Delta = 1$, while for $d=3$, we use $\Delta = 2$ and $\Delta = 1.5$. The former corresponds to the case where $\Delta >\frac{d}{2}$ while the latter has $\Delta = \frac{d}{2}$. These will give us two different boundary conditions for the scalar field. In the numerical calculations, we choose $L=1$.\\
\indent The relation between $r_0$ and $\Delta t$ for various value of $\tilde{\phi}_0=T^{-d+\Delta}\phi_0$ can be seen in figure \ref{fig:dt}, which recovers the analytical solution for $d=2$ and $\phi(r)=0$. The deformation parameter $\tilde{\phi}_0$ modifies the relation between $r_0$ and $\Delta t$ for hairy black hole solution. We also show that the relation between $r_0$ and $\Delta t$ in $d=2$ can be reasonably well approximated by the function
\begin{equation}\label{fittinginterval}
    \Delta t=2a\text{ tanh}^{-1}\bigg(\frac{(r_0-b)/a}{\sqrt{(c-(r_0-b)^2/a^2)}}\bigg),
\end{equation}
for some fitting parameters $\{a,b,c\}$ that can be obtained numerically. In this case, $b$ serves as a small regulator. Within this model, a new critical turning point $r=r_c$ where $\Delta t\rightarrow\infty$ emerges, which is given by
\begin{equation}
    r_c\approx a\sqrt{\frac{c}{2}},
\end{equation}
where we have ignore $b$. This critical turning point returns to BTZ case with $r_c=\frac{r_H}{\sqrt{2}}$ when $a=r_H,b=0,c=1$, and the relation between $r_0$ and $\Delta t$ returns to Eq. \eqref{dtanalytical}. With non-vanishing $\phi_0$, the value of $r_c$ shifts toward the horizon, increases to a maximum, and then decreases again. This shifted critical turning point is found to be larger than its BTZ counterpart. At small time interval, $\Delta t$ depends linearly on $r_0$, with the relation
\begin{equation}\label{deltatr0}
    \Delta t\approx\frac{2r_0}{\sqrt{c}},\;\;\;\Delta t\ll1.
\end{equation}
\indent To see the effect of the scalar deformation on the extremal surfaces, we examine the area density of both the spacelike and timelike surfaces. We regularize the result by subtracting the total area of $\gamma_\mathcal{T}$ with the area of the $K^2=0$ surface around $r=\varepsilon$, $\mathcal{A}(\gamma_0)$, or
\begin{equation}\label{renorarea}
    \text{Area}=\frac{\mathcal{A}(\gamma_\mathcal{T})-\mathcal{A}(\gamma_0)}{V_{d-2}L^{d-1}},
\end{equation}
where
\begin{equation}
    \frac{\mathcal{A}(\gamma_0)}{V_{d-2}L^{d-1}}=\int_\varepsilon^{r_H} \frac{dr}{r^{d-1}\sqrt{f}}.
\end{equation}
The area calculated in Eq. \eqref{renorarea} is the renormalized area density of the extremal surface $\gamma_\mathcal{T}$, which is now free of UV divergence and possibly negative. The result can be seen in figures \ref{fig:PlotRe} and \ref{fig:PlotIm} for the real ($\text{Re[Area]}$) and imaginary ($\text{Im[Area]}$) area, respectively. In $d=2$, our numerical results agree with the analytical BTZ calculations in the limit $\tilde{\phi}_0=0$.\\
\indent In both pure AdS$_3$ and BTZ cases, the imaginary part of the HTEE is a constant, which is given by $\text{Im[Area]}=\pi$, in agreement with the analytical result in Eq. \eqref{analyticalarea}. Interestingly, this no longer holds when we turn on the scalar deformation $\phi_0$ at the boundary. We find that, even in AdS$_3$, the imaginary part of the area depends on $\Delta t$, as shown in figure \ref{fig:PlotIm}. This result shows that the boundary deformation $\phi_0$ in the CFT, which corresponds to the bulk scalar field $\phi(r)$, significantly breaks the symmetry that timelike extremal surfaces have in pure AdS$_3$ and BTZ geometries. The scalar field backreacts on the bulk spacetime and induces a deformation throughout the entire geometry, which grows increasingly strong as we go deeper to the black hole interior. Since the timelike extremal surface extends deep into this interior region, it becomes highly sensitive to these large deformations, resulting in a modification of its area and symmetry structure.
\subsection{Analytical Near-Boundary Expansions at Small $\Delta t$}
\indent In $d=2$, we expect $\Im[\text{Area}] \to \pi$ as $\Delta t \to 0$ for any value of $\phi_0\geq0$. This can be analytically confirmed by taking the limit $r_0\rightarrow0$. In the limit $r_0 \to 0$, the dominant contribution to the total area arises from integrations performed in the near-boundary region. Therefore, we could expand the function $f(r)$ and $\chi(r)$ around $r\rightarrow0$, inserting this expansion to the equations of motion, and obtain the expansion coefficients. The near-boundary expansions are given by (see, for example, \cite{gao2024,Prihadi2025b})
\begin{align}
\phi(r)&=\phi_0r^{d-\Delta}+\frac{\langle\mathcal{O}\rangle}{2\Delta -d}r^\Delta +...,\\
\chi(r)&=\frac{d-\Delta}{2(d-1)}\phi_0^2 r^{2(d-\Delta)}+...,\\ fe^{-\chi}&=1-\langle T_{tt}\rangle r^{d}+..., \label{fchiexpansion}
\end{align}
where $\langle T_{tt}\rangle$ is the expectation value of the energy-momentum tensor in the corresponding CFT theory, which can also be expressed in terms of $\phi_0$ and $\langle\mathcal{O}\rangle$.\\
\indent The area functionals can now be written as
\begin{align}
    \frac{\mathcal{A}_+(\gamma_\mathcal{T})}{V_{d-2}L^{d-1}}=&2\int_{\varepsilon}^{r_\star}\frac{dr}{r^{d-1}}\frac{1}{\sqrt{f}}\frac{1}{\sqrt{1+(r/r_0)^{(2d-2)}\frac{f_0e^{-\chi_0}}{fe^{-\chi}}}}\\\nonumber
    \approx&2\int_\varepsilon^{r_\star}\frac{dr}{r^{d-1}}\frac{e^{-\chi/2}}{\sqrt{1+(r/r_0)^{2d-2}}},
\end{align}
for the spacelike surfaces, and
\begin{align}
    \frac{\mathcal{A}_-(\gamma_\mathcal{T})}{V_{d-2}L^{d-1}}=&2i\int_{r_0}^{r_\star}\frac{dr}{r^{d-1}}\frac{1}{\sqrt{f}}\frac{1}{\sqrt{(r/r_0)^{(2d-2)}\frac{f_0e^{-\chi_0}}{fe^{-\chi}}-1}}\\\nonumber
    \approx&2i\int_{r_0}^{r_\star}\frac{dr}{r^{d-1}}\frac{e^{-\chi/2}}{\sqrt{(r/r_0)^{2d-2}-1}},
\end{align}
for the timelike surfaces, where we have used the near-boundary expansions, ignoring all of the subleading terms in Eq. \eqref{fchiexpansion}. We introduce an intermediate cutoff $r_\star$, with $1 > r_\star > r_0>\varepsilon$, chosen such that the contribution from the region $r_\star < r < \infty$ can approximately be ignored. In this regime, the integral is dominated by the near-boundary region while the asymptotic expansion remains valid. Although $r_\star$ is not physical, it serves to parametrize our ignorance of the IR completion of the merged extremal surfaces.\\
\indent The factor $e^{-\chi/2}$ can also be expanded in small $r$, leading to
\begin{equation}\label{chiexpansion}
    e^{-\chi/2}=\sum_{n=0}^\infty \alpha_n(\phi_0,\langle\mathcal{O}\rangle)r^{2n(d-\Delta)},
\end{equation}
where $\alpha_n(\phi_0,\langle\mathcal{O}\rangle)$ is the expansion coefficient for each $n$, which depends on the boundary data $\phi_0$ and $\langle\mathcal{O}\rangle$. At the zeroth order, we have $\alpha_0=1$ and the first order gives us $\alpha_1=\frac{(d-\Delta)\phi_0^2}{4(d-1)}$. Using this expansion, the area functionals now become
\begin{align}\label{nearvacuumareareal}
    &\frac{\mathcal{A}_+(\gamma_\mathcal{T})}{V_{d-2}L^{d-1}}=2\sum_{n=0}^\infty\alpha_n\int_{\varepsilon}^{r_\star}\frac{dr}{r^{d-1}}\frac{r^{2n(d-\Delta)}}{\sqrt{1+(r/r_0)^{2d-2}}},
\end{align}
and
\begin{align}\label{nearvacuumareaimaginary}
&\frac{\mathcal{A}_-(\gamma_\mathcal{T})}{V_{d-2}L^{d-1}}=2i\sum_{n=0}^\infty\alpha_n\int_{r_0}^{r_\star}\frac{dr}{r^{d-1}}\frac{r^{2n(d-\Delta)}}{\sqrt{(r/r_0)^{2d-2}-1}}.
\end{align}
The integrals can be solved using the hypergeometric functions ${}_2F_1(a,b,c;z)$. In the $\Delta t \to 0$ limit, the turning point $r_0$ is far from the horizon. As a result, the $n=0$ term reproduces the timelike entanglement entropy of a strip $\mathcal{T}$ in vacuum AdS$_{d+1}$, while the $n>0$ contributions correspond to near-vacuum CFT corrections, since the extremal surface probes only the asymptotically AdS region and is insensitive to the black hole geometry.\\
\indent It is instructive to consider an example with $(d,\Delta)=(2,1.5)$. In this background, the area functional with the first-order correction from the factor $e^{-\chi/2}$ can be computed as follows:
\small
\begin{align}\label{fitred2D32}
    &\frac{\mathcal{A}_+(\gamma_\mathcal{T})}{V_{d-2}L^{d-1}}=2\int_\varepsilon^{r_\star}\frac{dr}{r}\frac{1-(\frac{1}{2}-\frac{\Delta}{4})\phi_0^2r^{2(2-\Delta)}}{\sqrt{1+(r/r_0)^2}}\\\nonumber
    &\quad=2\ln\bigg(\frac{2r_0}{\varepsilon}\bigg)-2\text{arcsinh}\bigg(\frac{r_0}{r_\star}\bigg)\\\nonumber
    &\quad+\frac{\phi_0^2}{2}\frac{\Delta -2}{3-2\Delta}r_0r_\star^{3-2\Delta}{}_2F_1\bigg(\frac{1}{2},\Delta-\frac{3}{2},\Delta-\frac{1}{2};\frac{r_0^2}{r_\star^2}\bigg)\\\nonumber
    &\quad\approx2\ln\bigg(\frac{2r_0}{\varepsilon}\bigg)+\frac{\phi_0^2}{4}r_0\ln\bigg(\frac{r_0}{2r_\star}\bigg)-\frac{\phi_0^2}{4}\frac{r_0^3}{4r_\star^2}+\dots\;,
\end{align}
\normalsize
for the spacelike surface, and
\small
\begin{align}\label{fitd2D32}
    &\frac{\mathcal{A}_-(\gamma_\mathcal{T})}{V_{d-2}L^{d-1}}=2i \int_{r_0}^{r_\star} \frac{dr}{r} \frac{1 - (\frac{1}{2}-\frac{\Delta}{4}) \phi_0^2 r^{2(2-\Delta)}}{\sqrt{(r/r_0)^2 - 1}} \nonumber\\ &\;\;\;\;\;= 2i \arccos\left(\frac{r_0}{r_\star}\right) \nonumber \\
&\;\;\;\;\;\quad + \frac{i \sqrt{\pi} \phi_0^2}{4} r_0^{4-2\Delta} \frac{\Gamma(\Delta - \frac{3}{2})}{\Gamma(\Delta - 2)} \nonumber \\
&\;\;\;\;\;\quad + \frac{i \phi_0^2}{2} \frac{\Delta - 2}{3 - 2\Delta} r_0 r_\star^{3-2\Delta} \, _2F_1\left( \frac{1}{2}, \Delta - \frac{3}{2}, \Delta - \frac{1}{2}; \frac{r_0^2}{r_\star^2} \right) \nonumber\\ &\;\;\;\;\;\approx i\pi + \frac{i \phi_0^2}{4}r_0\ln\left(\frac{r_0}{2 r_\star}\right)+i\frac{\phi_0^2}{4}\frac{r_0^3}{4r_\star^2}+\dots\;,
\end{align}
\normalsize
for the timelike surface, where the approximation is taken by expanding around $\Delta = 1.5$ and assuming small $r_0$. In figure \ref{fig:PlotIm}, we fit the analytical approximation obtained in the small–$\Delta t$ regime to the numerical data and extract an approximate numerical value for $r_\star$. This fit is performed using the relation in Eq. \eqref{deltatr0}, which connects the turning point to the boundary time interval. As an example, we find that $r_\star=0.50537$ when $\tilde{\phi}_0=5.0375$, which is in the range of $1>r_\star>r_0$. The quality of the fit improves for smaller values of $\phi_0^2/4$, consistent with the regime of validity of the perturbative expansion. For the case where $(d,\Delta)=(2,1)$, the expression yields a power law correction, as shown below:
\begin{align}
    \frac{\mathcal{A}_+(\gamma_\mathcal{T})}{V_{d-2}L^{d-1}}\approx2\ln\bigg(\frac{2r_0}{\varepsilon}\bigg)-\frac{\phi_0^2}{2}r_0r_\star+\frac{\phi_0^2}{4r_\star}r_0^3+\dots\;,
\end{align}
and
\begin{align}\label{fitd2D1}
\frac{\mathcal{A}_-(\gamma_\mathcal{T})}{V_{d-2}L^{d-1}} &\approx  i\pi - \frac{i\phi_0^2}{2}r_0r_\star-\frac{i\phi_0^2}{4r_\star}r_0^3+\dots\;.
\end{align}
\indent Likewise, for $(d,\Delta)=(3,2)$, the expansion retains the standard $1/r_0$ divergence, with the scalar deformation appearing as a subleading linear correction term proportional to $r_0$:
\begin{align}\label{fitd3D2}
    \frac{\mathcal{A}_+(\gamma_\mathcal{T})}{V_{d-2}L^{d-1}}\approx&\frac{2}{\varepsilon}-\frac{2\sqrt{\pi}\Gamma(\frac{3}{4})}{r_0\Gamma(\frac{1}{4})}-\frac{\phi_0^2\sqrt{\pi}\Gamma(\frac{1}{4})}{16\Gamma(\frac{3}{4})}r_0\\\nonumber
    &+\frac{\phi_0^2}{4r_\star}r_0^2+\dots\;,
\end{align}
and
\begin{align}
\frac{\mathcal{A}_-(\gamma_\mathcal{T})}{V_{d-2}L^{d-1}} &\approx \frac{i}{r_0} \frac{2\sqrt{\pi} \Gamma(\frac{3}{4})}{\Gamma(\frac{1}{4})} - \frac{i\phi_0^2 \sqrt{\pi} \Gamma(\frac{1}{4})}{16 \Gamma(\frac{3}{4})} r_0\nonumber\\ &\quad+\frac{i\phi_0^2}{4r_\star}r_0^2+\dots\;.
\end{align}
Finally, for $(d,\Delta)=(3,1.5)$, the area functional with the first-order correction can be evaluated as follows:
\begin{align}\label{fitd3D32}
   \frac{\mathcal{A}_+(\gamma_\mathcal{T})}{V_{d-2}L^{d-1}} \approx&\frac{2}{\varepsilon}-\frac{2\sqrt{\pi}\Gamma(\frac{3}{4})}{r_0\Gamma(\frac{1}{4})}+\frac{3\phi_0^2}{8}r_0^2\ln\bigg(\frac{r_0}{r_\star\sqrt{2}}\bigg)+\dots\;,
\end{align}
and
\begin{align}
    \frac{\mathcal{A}_-(\gamma_\mathcal{T})}{V_{d-2}L^{d-1}} &\approx \frac{i}{r_0}  \frac{2\sqrt{\pi}\Gamma(\frac{3}{4})}{\Gamma(\frac{1}{4})} +\frac{3 i \phi_0^2}{8} r_0^2 \ln\left(\frac{r_0}{r_\star \sqrt{2}}\right)+\dots\;.
\end{align} 
In this case, the leading $\sim\frac{i}{r_0}$ behavior gets corrected by $\sim r_0^2\ln r_0$ when the scalar field is turned on at $\Delta t\ll1$. We find an agreement between numerical results and analytical calculations for both $d =2$ and $d=3$ in small $\Delta t$ limit. When the scalar field is turned off, all expressions reduce to those obtained in the vacuum case found in \cite{Doi2023}, using the relation
\normalsize
\begin{equation}
    r_0=\frac{\Gamma(\frac{1}{2(d-1)})}{2\sqrt{\pi}\Gamma(\frac{d}{2(d-1)})}\Delta t.
\end{equation}
Moreover, $r_0\rightarrow0$ limit also reproduces the vacuum case as the scalar field does not contribute in this limit. This is physically expected, since $r_0$ controls how deeply the extremal surface probes the bulk geometry and the effect of the scalar field $\phi(r)$ is suppressed near the asymptotic boundary as the deformation is relevant.\\
 \subsection{Comparison with Temporal Entanglement and Spacelike Entanglement}
 In Ref. \cite{Doi2023}, the timelike entanglement entropy is defined by analytically continuing the (spacelike) entanglement entropy of a spatial subsystem of size $\Delta x$ in a quantum many-body system to the timelike subsystem with interval $\Delta t$ via a Wick rotation, $\Delta x \rightarrow i\,\Delta t$. In this section, we calculate the spacelike entanglement entropy and the temporal entanglement entropy in the hairy black hole background. The latter is obtained by first performing a Wick rotation to the time coordinate $t\rightarrow i \tau_E$, so that the metric becomes Euclidean,
 \begin{equation}
     ds_E^2=\frac{L^2}{r^2}\bigg(fe^{-\chi}d\tau_E^2+\frac{dr^2}{f}+dx^2+d\textbf{y}_{d-2}^2\bigg).
 \end{equation}
We calculate the spacelike entanglement entropy for a strip-like boundary subregion $A$ with length $\Delta x$ and the temporal entanglement entropy for a temporal strip with a Euclidean time interval $\Delta \tau_E$.\\
\indent The temporal entanglement entropy is calculated by the area given by
\begin{equation}
    \frac{\mathcal{A}(\gamma_{\mathcal{\tau}_E})}{V_{d-2}L^{d-1}}=2\int_\varepsilon^{r_0}\frac{dr}{r^{d-1}}\sqrt{fe^{-\chi}\tau_E^{\prime 2}+\frac{1}{f}},
\end{equation}
with
\begin{equation}
    \tau_E^\prime(r)^2=\frac{f_0e^{-\chi_0}}{fe^{-\chi}}\frac{(r/r_0)^{2d-2}}{fe^{-\chi}-(r/r_0)^{2d-2}f_0e^{-\chi_0}}.
\end{equation}
In the near-vacuum case with $fe^{-\chi}\approx1$, the temporal entanglement entropy is calculated by
\begin{align}
    \frac{\mathcal{A}(\gamma_{\mathcal{\tau}_E})}{V_{d-2}L^{d-1}}=&2\int_\varepsilon^{r_0}\frac{dr}{r^{d-1}}\frac{e^{-\chi/2}}{\sqrt{1-(r/r_0)^{2d-2}}}\\\nonumber
    =&2\sum_{n=0}^\infty\alpha_n\int_\varepsilon^{r_0}\frac{dr}{r^{d-1}}\frac{r^{2n(d-\Delta)}}{\sqrt{1-(r/r_0)^{2d-2}}}.
\end{align}
This integral, although similar, is markedly different compared with the HTEE calculations in Eqs. \eqref{nearvacuumareareal} and \eqref{nearvacuumareaimaginary}. The main difference lies in the upper and lower limit of the integral. In the temporal entanglement case, the integral is performed from $\varepsilon$ to $r_0$, where $\varepsilon<r_0$ is always satisfied. This could modify the hypergeometric functions and result in a different logarithmic behavior in the expansions. This difference reflects the fact that temporal entanglement entropy only probes the region between the UV boundary and the turning point, whereas the HTEE necessarily contains additional bulk regions beyond the turning point, presented in $r_\star>r_0$.\\
\indent For example, in the case with $(d,\Delta)=(2,1.5)$, we obtain
\small
\begin{align}
    \frac{\mathcal{A}(\gamma_{\mathcal{\tau}_E})}{L}=&2\int_\varepsilon^{r_0} \frac{dr}{r}\frac{1-(\frac{1}{2}-\frac{\Delta}{4})\phi_0^2r^{2(2-\Delta)}}{\sqrt{1-(r/r_0)^2}}\\\nonumber
    =&2\ln\bigg(\frac{2r_0}{\varepsilon}\bigg)-\frac{\phi_0^2}{4}\frac{\sqrt{\pi}r_0^{2(2-\Delta)}\Gamma(3-\Delta)}{\Gamma\big(\frac{5}{2}-\Delta)}\\\nonumber
    &\quad+\frac{\phi_0^2}{4}\varepsilon^{2(2-\Delta)}{}_2F_1\bigg(\frac{1}{2},2-\Delta,3-\Delta;\frac{\varepsilon^2}{r_0^2}\bigg)\\\nonumber\approx&2\ln\bigg(\frac{2r_0}{\varepsilon}\bigg)-\frac{\phi_0^2}{4}\frac{\pi r_0}{2}+\dots\;,
\end{align}
\normalsize
upon taking the $\varepsilon\rightarrow0$ limit, and expanding near $r_0=0$ while keeping $r_0>\varepsilon$. Here, we find a correction term linear to $r_0$, without the $\sim r_0\ln r_0$ term. The turning point $r_0$ is linearly proportional to $\Delta \tau_E$ for small $r_0$. Therefore, identifying $r_0\rightarrow ir_0$ implies an analytic continuation from $\Delta\tau_E\rightarrow i\Delta t$.\\
\indent One could see that, by taking the analytic continuation of the temporal entanglement for $(d,\Delta)=(2,1.5)$, we do not recover the HTEE obtained in Eqs. \eqref{fitred2D32} and \eqref{fitd2D32}. However, by reversing the analytic continuation of the HTEE from identifying $r_0\rightarrow -i r_0$ in Eqs. \eqref{fitred2D32} and \eqref{fitd2D32}, we obtain
\begin{align}\label{reverseanalyticd2D32}
    2\ln\bigg(\frac{2r_0}{\varepsilon}\bigg)-\frac{\phi_0^2}{4}\frac{\pi r_0}{2}+\dots+i\frac{\phi_0^2}{4}\frac{\pi r_0}{2}\;+\mathcal{O}(r_0^2),
\end{align}
which approximately reproduces $\mathcal{A}(\gamma_{\mathcal{\tau}_E})$ with an additional imaginary term. The $\dots$ in Eq. \eqref{reverseanalyticd2D32} contain both real and imaginary terms that depend explicitly on the intermediate radial cutoff $r_\star$. We argue that the discrepancies contained in the $\dots$ terms are due to our limitation that only allows us to probe the near-boundary region and enforce an intermediate cutoff beyond $r_\star$. Such terms are sensitive to the IR completion of both spacelike and timelike extremal surfaces in the deep-bulk region, or even in the black hole interior, where those surfaces are merged. On the other hand, the temporal entanglement entropy is not expected to depend on $r_\star$, since the corresponding extremal surface extends only from the UV cutoff to the turning point $r_0$, which remains smaller than $r_\star$. This gives the holographic prescription a nontrivial consistency check at the level of the near-boundary expansions. At present, whether the full analytic solutions coincide under analytic continuation remains an open question, as the complete analytic solution in the hairy black hole background is required.\\
\indent In $d=3$, the temporal entanglement entropy is given by
\begin{align}
    \frac{\mathcal{A}(\gamma_{\mathcal{\tau}_E})}{V_{1}L^{2}}=&\frac{2}{\varepsilon}-\frac{2\sqrt{\pi}\Gamma(\frac{3}{4})}{r_0\Gamma(\frac{1}{4})}-\frac{\phi_0^2}{4}\frac{r_0\sqrt{\pi}\Gamma(\frac{1}{4})}{4\Gamma(\frac{3}{4})}+\dots\;,
\end{align}
for $\Delta =2$, and
\begin{align}
    \frac{\mathcal{A}(\gamma_{\mathcal{\tau}_E})}{V_{1}L^{2}}=&\frac{2}{\varepsilon}-\frac{2\sqrt{\pi}\Gamma(\frac{3}{4})}{r_0\Gamma(\frac{1}{4})}-\frac{3\phi_0^2}{4}\frac{\pi r_0^2}{8}+\dots\;,
\end{align}
for $\Delta=1.5$. Here, we find similar $r_0$ scaling behavior between temporal entanglement entropy and HTEE in $(d,\Delta)=(3,2)$ for the term that does not depends on $r_\star$, while the $\sim r_0^2$ term in the $(d,\Delta)=(3,1.5)$ case can be obtained from taking the reverse analytic continuation $r_0\rightarrow-i r_0$ on Eq. \eqref{fitd3D2}.\\
\indent We therefore remarkably find that, in small $\Delta t$ limit, once the relevant boundary deformation is turned on, the analytic continuation of the CFT entanglement entropy does not exactly reproduce the HTEE calculations. This mismatch persists even in $d=2$, indicating that the deformation breaks the analytic equivalence between spacelike and timelike entanglement that exists in the undeformed theories. This demonstrates that the HTEE captures genuinely new information that cannot be inferred from spacelike observables once a  relevant deformations are present.\\
\indent For the spacelike entanglement entropy, the area functional is given by
\begin{equation}
    \frac{\mathcal{A}(\gamma_A)}{V_{d-2}L^{d-1}}=2\int\frac{dr}{r^{d-1}}\sqrt{x'^2+\frac{1}{f}},
\end{equation}
with the extremal surface $x(r)$ satisfying the Euler-Lagrange equation
\begin{equation}
    x^\prime(r)^2=\frac{r^{2d-2}}{r_0^{2d-2}}\frac{1}{f(1-(r/r_0)^{2d-2})}.
\end{equation}
The area then becomes
\begin{align}
    \frac{\mathcal{A}(\gamma_A)}{V_{d-2}L^{d-1}}=2\int_\varepsilon^{r_0}\frac{dr}{r^{d-1}}\frac{1}{\sqrt{f}}\frac{1}{\sqrt{1-(r/r_0)^{2d-2}}}.
\end{align}
At the leading order, the function $f(r)$ can be expanded as 
\begin{equation}
    f(r)\approx1+\frac{d-\Delta}{2(d-1)}\phi_0^2r^{2(d-\Delta)}+...\;.
\end{equation}
Consequently, both spacelike and temporal entanglement entropy coincide at leading order of the near-boundary expansion, and the same analysis underlying the analytic continuation of the holographic timelike entanglement entropy explained earlier applies equally well to the spacelike case. Deviations arise only from IR contributions beyond the regime of validity of the asymptotic expansion.
\subsection{Late-Time Behavior and Exterior vs. Interior Contributions}
\begin{figure}
     \centering
     \includegraphics[width=0.47\linewidth]{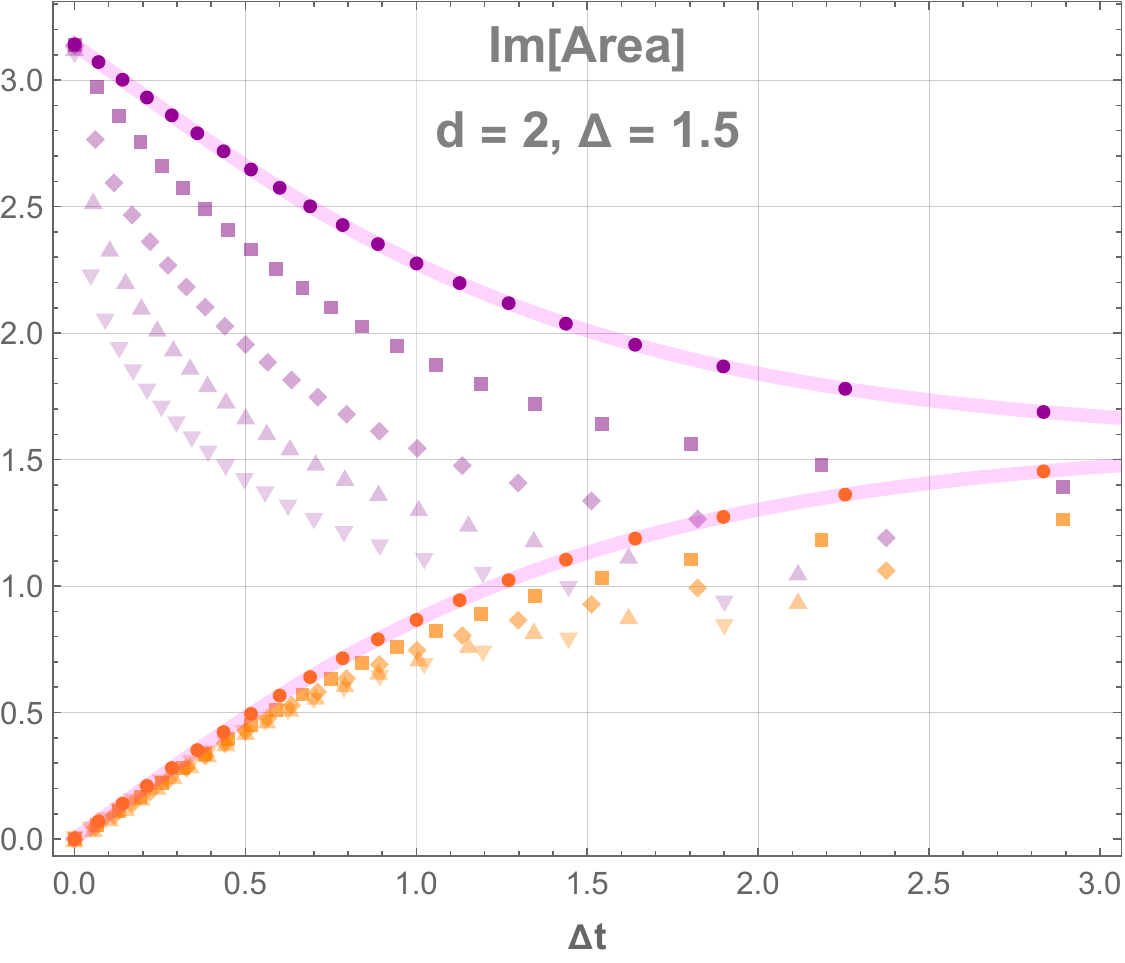}
     \includegraphics[width=0.47\linewidth]{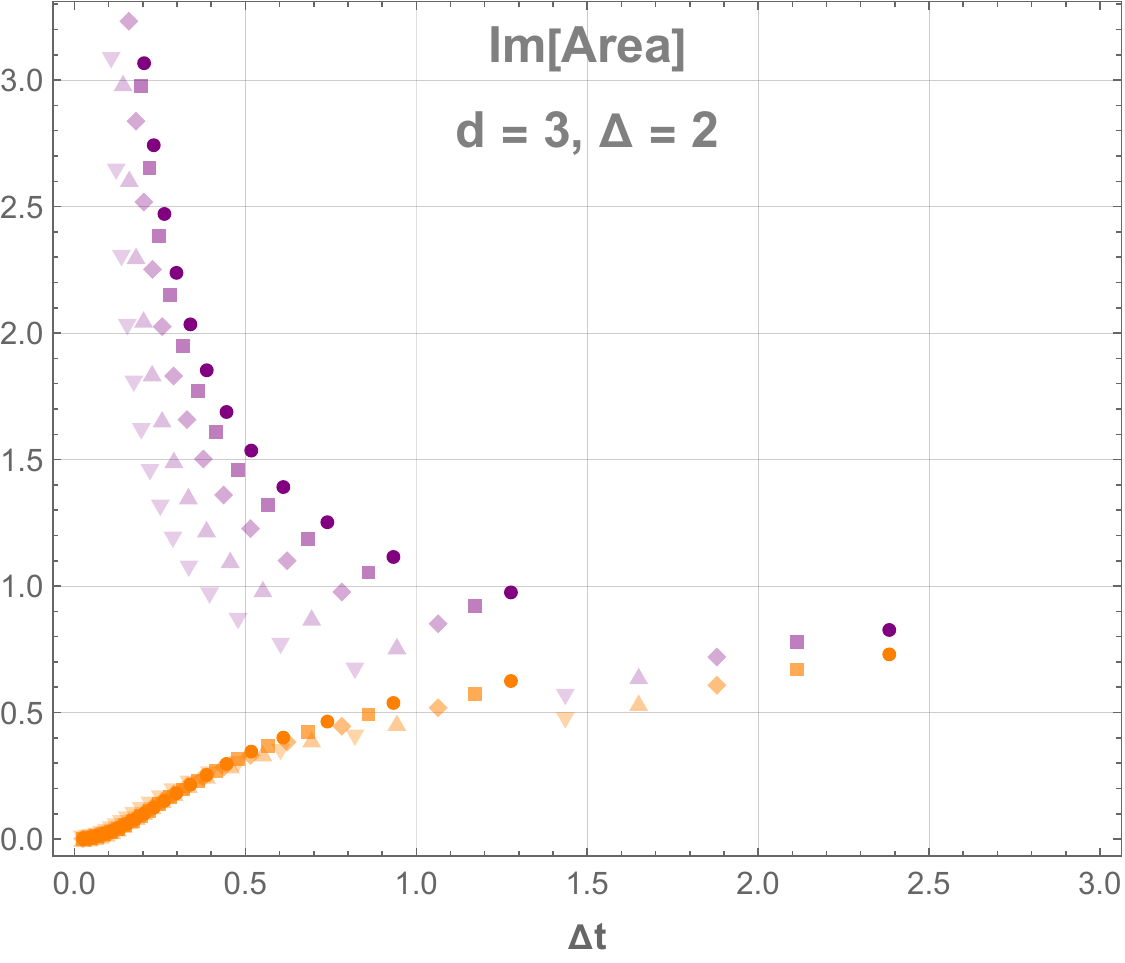}\\
     \includegraphics[width=0.47\linewidth]{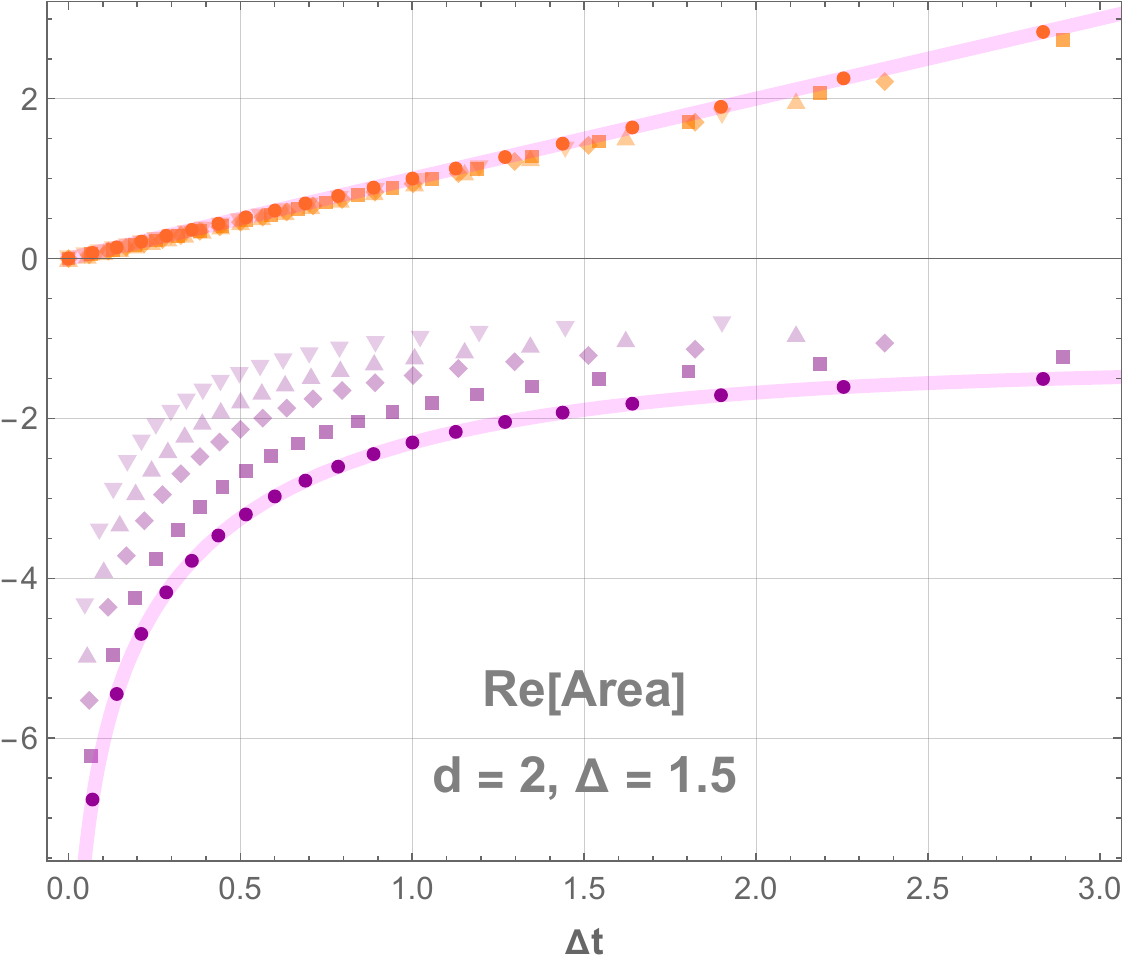}
     \includegraphics[width=0.47\linewidth]{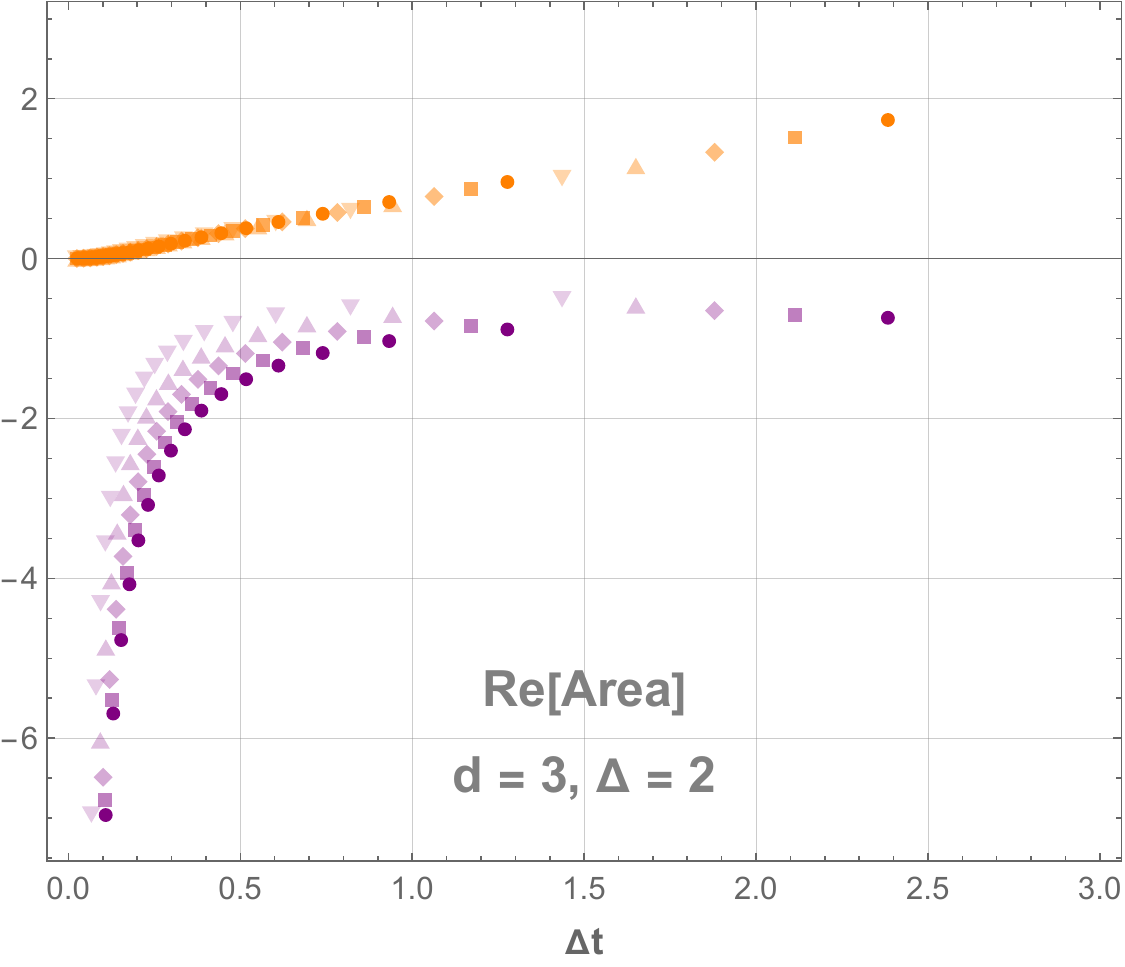}
     \caption{Exterior (purple) vs. interior (orange) contributions to the area. Solid magenta lines represent the known analytical solution. Decreasing opacity depicts increasing $\tilde{\phi_0}$.}
     \label{fig:extint}
 \end{figure}

Previous analyses discuss the analytical approach to the small $\Delta t$ approximations as the turning point is far from the presence of the horizon. At large $\Delta t$, however, the existence of the horizon cannot be ignored anymore. Here, the analytical approach becomes subtle since we need to solve the complete equations of motion for $f(r)$ and $\chi(r)$. In this subsection, we present late-time analysis based on our numerical results.\\
\indent At large $\Delta t$, $\Im[\text{Area}]$ scales linearly in $\Delta t$ for $d=2$. Similar late-time behavior also occurs in higher dimensions. In $d=3$, $\Im[\text{Area}]$ behaves as
\begin{equation}\label{ImaginaryFit1}\Im[\text{Area}]\sim\frac{a}{\Delta t}+c,\;\;\;\text{as}\;\;\;\Delta t\ll1,
\end{equation}
or
\begin{equation}
    \label{ImaginaryFit2}
    \Im[\text{Area}]\sim b\Delta t+d,\;\;\;\text{as}\;\;\;\Delta t\gg1.
\end{equation}
We fit those functions to the numerical results to get the fitting parameters $\{a,b,c,d\}$ for various $\tilde{\phi}_0$. On the other hand, $\Re[\text{Area}]$ scales linearly with $\Delta t$ as $\Delta t\gg1$.
 Generally, in any dimension, the scalar field $\tilde{\phi}_0$ increases the real part of the total area while it decreases the imaginary part.\\
 \indent One can also see how the extremal surfaces in the exterior and the interior of the black hole contribute to the calculation of the total area. As $\Delta t\rightarrow\infty$, the turning point becomes closer to the horizon and the surface area spends an increasingly large portion in the black hole interior. In this regime, the dominant contribution to $\Re[\text{Area}]$ arises from the interior segment, while the exterior contribution saturates. This can be figuratively confirmed in figure \ref{fig:extint}. The linear growth of $\Re[\text{Area}]$ at late time is governed by the interior solution.\\
 \indent In contrast, $\Im[\text{Area}]$ receives comparable contributions from both the exterior and interior segments, which become symmetric at large $\Delta t$. The symmetry between the interior and exterior contributions to $\Im[\text{Area}]$ at late times reflects the balanced analytic continuation structure of the extremal surfaces. In AdS$_3$ with vanishing $\phi(r)=0$, $\Im[\text{Area}]$ is always symmetric between its exterior and interior contributions, and the total imaginary part is always independent of $\Delta t$. The presence of a scalar hair breaks this exterior–interior symmetry, leading to a nontrivial $\Delta t$-dependence imaginary part even in AdS$_3$. Taken together, these observations indicate that the late-time behavior of the HTEE is predominantly governed by the black hole interior, while the imaginary part becomes sensitive to the contributions between interior and exterior regions induced by the scalar deformation.
\subsection{Remarks on Charged Hairy Black Holes}
\begin{figure}
    \centering
    \includegraphics[width=0.45\linewidth]{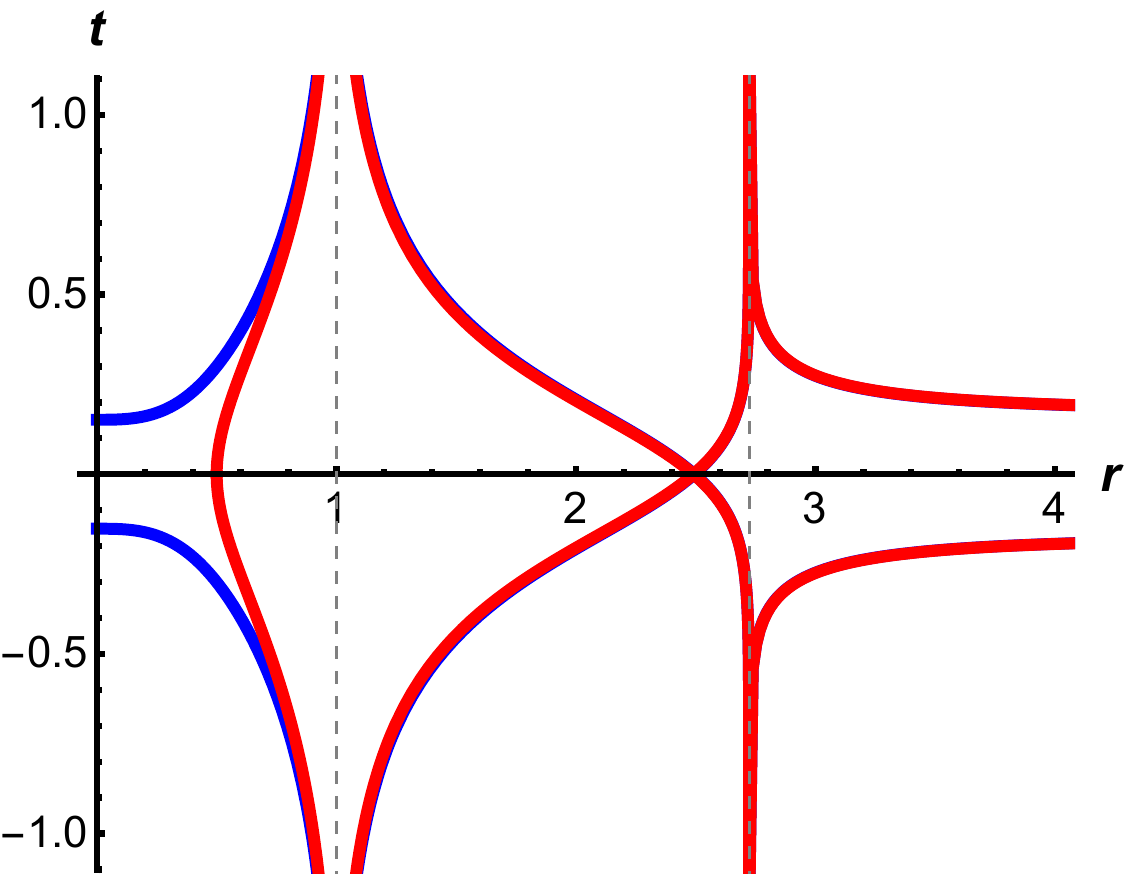}\;
    \includegraphics[width=0.45\linewidth]{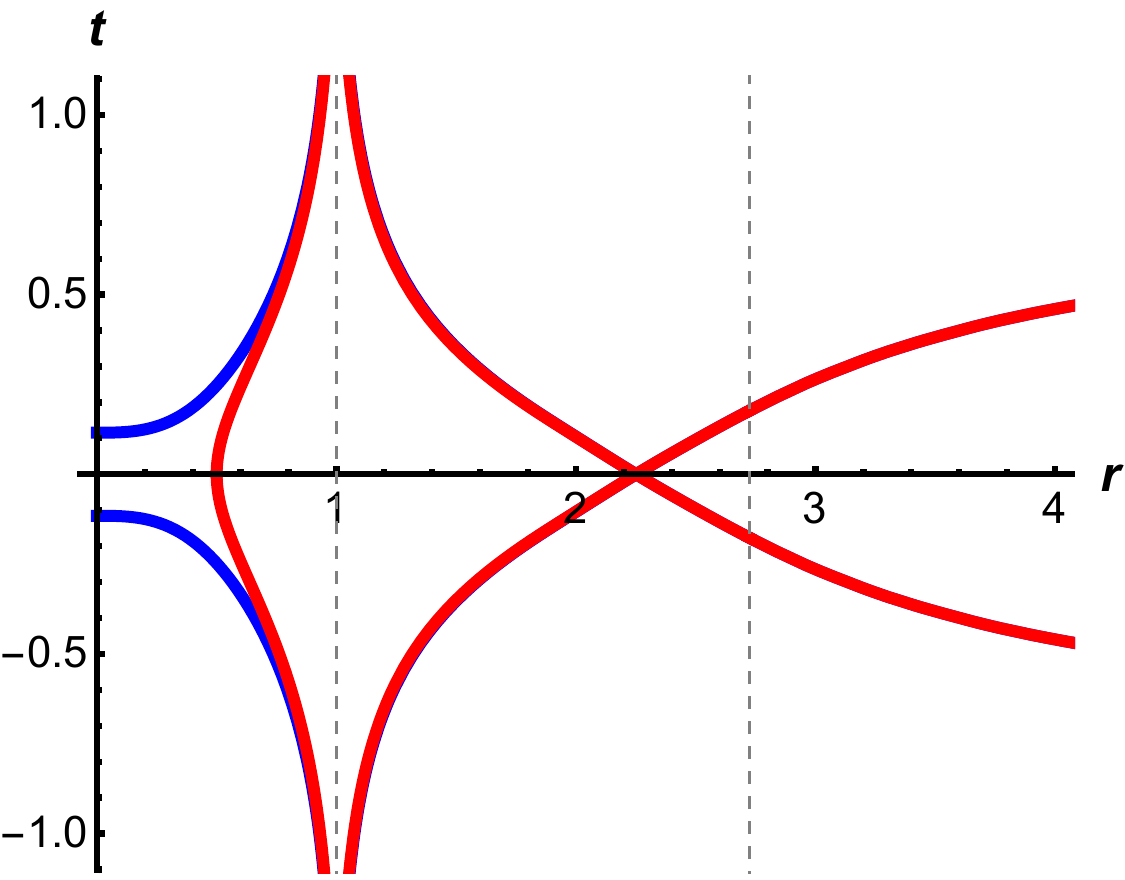}
    \caption{Illustration of the extremal surfaces in charged black hole with $\tilde{\phi}_0=0$ (left) and $\tilde{\phi}_0>0$ (right). The dashed vertical line at $r=1$ represents the outher horizon while $r\approx2.723$ is the would-be inner horizon when $\phi(r)=0$.}
    \label{fig:HTEEchagred}
\end{figure}
Both $t_+$ ($t^\prime_+$) and $t_-$ ($t^\prime_-$) goes to $\infty$ at the horizon due to the factor of $f(r)$ in Eq. \eqref{tprimes}. Therefore, if the black hole has multiple horizons, we expect the same divergence to occur at each horizon. For example, a charged Reissner-Nordstr\"om-AdS black hole could have an outer $r_+$ and inner $r_-$ horizon. In this case, we identify $t_s(r_\pm-\delta)=t_s(r_\pm+\delta)$ for both $s=\pm1$. The spacelike and timelike surfaces are then merged at $r=\infty$ inside the inner horizon. After this merger, one obtains a single continuous surface homologous to $\mathcal{T}$, as illustrated in figure \ref{fig:HTEEchagred}.\\
\indent However, once the boundary relevant deformation $\phi_0$ is turned on, the inner horizon is destroyed, and the interior structure dramatically changes into an Einstein-Rosen bridge collapse right after the would-be inner horizon \cite{Hartnoll2020}. In this case, the extremal surface $\gamma_\mathcal{T}$ undergoes a transition, shifting from the configuration shown in the left panel to that in the right panel of figure \ref{fig:HTEEchagred}. However, this transition does not affect the behavior of $(t^\prime_s)^2$ at $r\rightarrow\infty$. This is because, charged hairy black holes also have similar Kasner behavior as in Eq \eqref{fieldskasner}. Therefore, the spacelike-timelike extremal surfaces can still be consistently joined at $r\rightarrow\infty$.\\
\indent Another important aspect here is that the extremal surfaces could possibly intersect each other in the interior region between $r_+$ and $r_-$. This would violate the entanglement wedge nesting property discussed in \cite{Czech2012,Wall2014,akers2016}. This crossing can be avoided by choosing different branches of the extremal surfaces since Eq \eqref{tprimesquare} gives us two (positive and negative) solutions. In Eq. \eqref{tprimes}, we automatically choose the positive branch. By choosing the positive $t^\prime_s(r)$ branch in the exterior $0<r<r_+$ and choosing the negative ones in the interior $(r>r_+$), we could obtain a non-crossing smooth extremal surface $\gamma_\mathcal{T}$. This also applies to the case when the scalar field is turned on, as we observe in figure \ref{fig:HTEEchagred} that the extremal surfaces also cross in the region between the outer horizon and the would-be inner horizon.\\
\indent A different choice of the $t^\prime_s(r)$ branch does not change the area of the minimal surface, as it only depends on $t^\prime_s(r)^2$. However, interestingly, this would affect the subregion complexity, as $\Delta V$ depends linearly on $t_s^\prime(r)$. This will be elaborated in the next section.
\section{Timelike Holographic Subregion Complexity}\label{sec4}
\begin{figure}
    \centering    \includegraphics[width=0.45\linewidth]{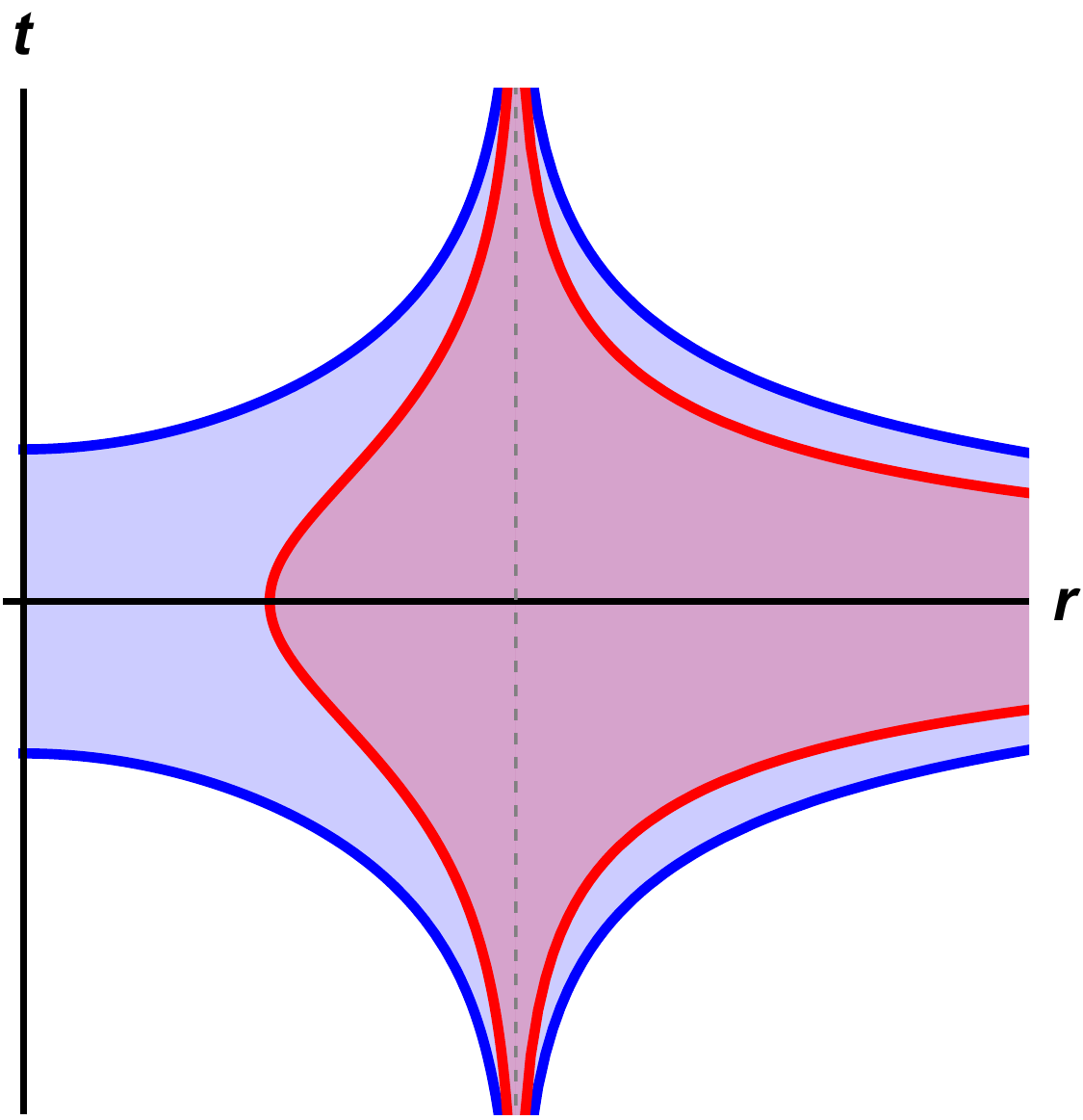}\;\;\;\;\;\;\;\;
\includegraphics[width=0.45\linewidth]{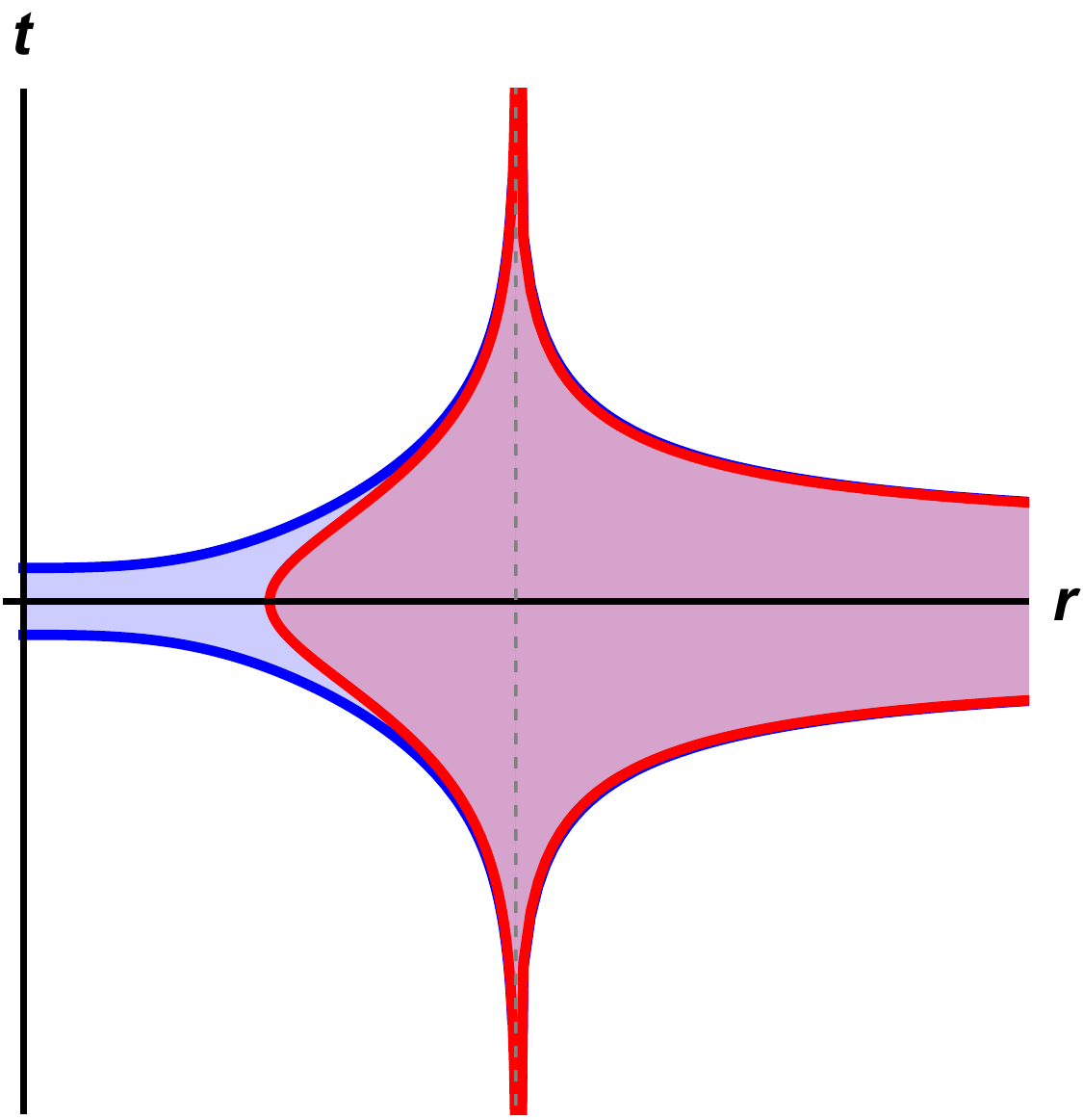}
    \caption{Illustration of the volume enclosed by $t_+$ surfaces (blue) and $t_-$ surfaces (red) for $\tilde{\phi}_0=0$ (left) and $\tilde{\phi}_0=2.1533$ (right). Here we choose $d = 2$, $\Delta =1.5$, $r_H=1$, and $r_0=0.5$.}
    \label{fig:CV}
\end{figure}
\begin{figure*}
    \centering
    \includegraphics[width=0.45\linewidth]{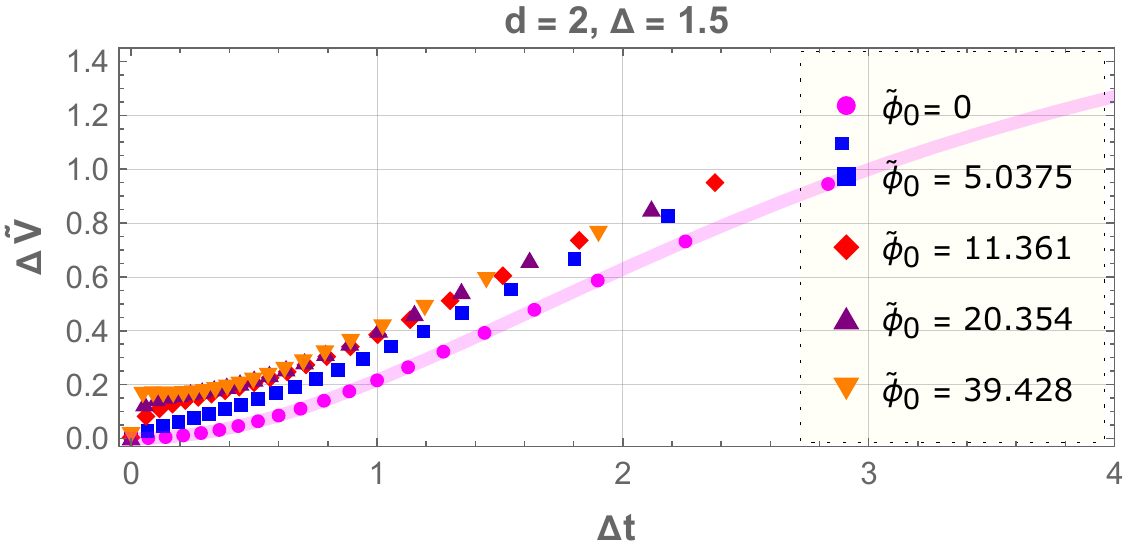}\;\;\;\;\;
    \includegraphics[width=0.45\linewidth]{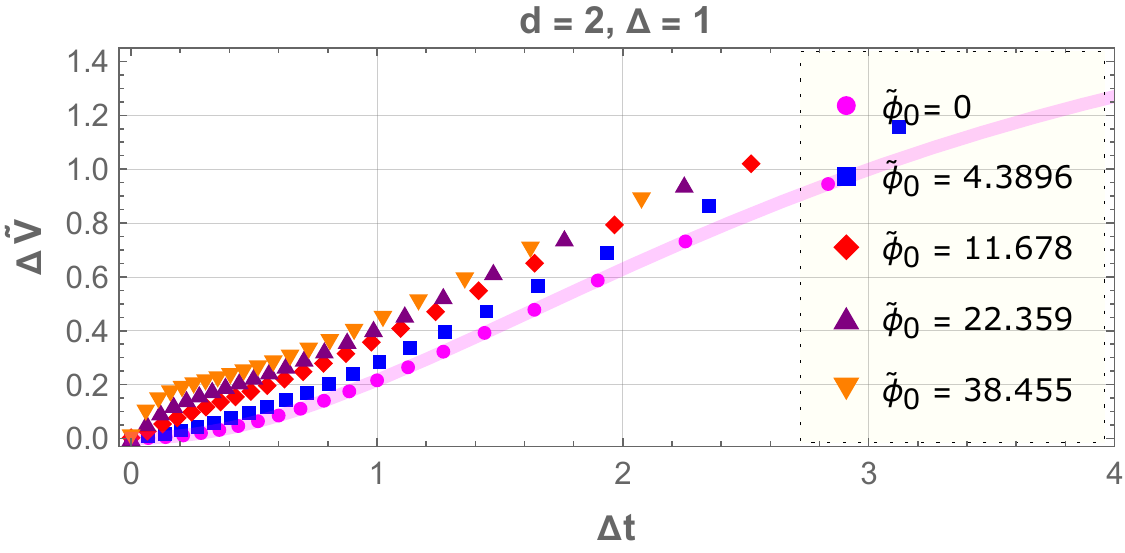}\\
    \includegraphics[width=0.45\linewidth]{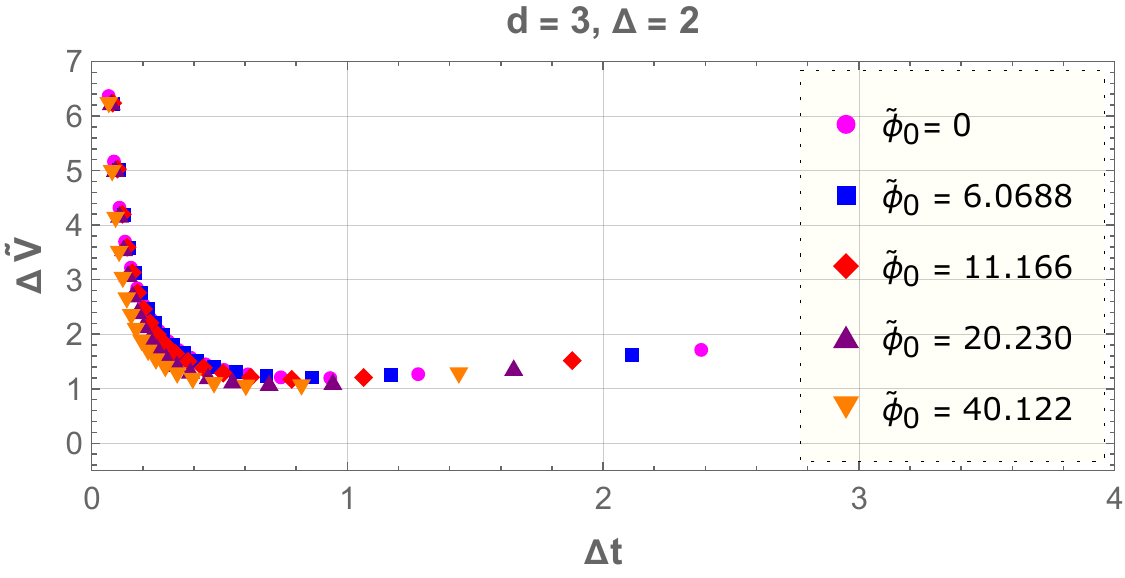}\;\;\;\;\;
    \includegraphics[width=0.45\linewidth]{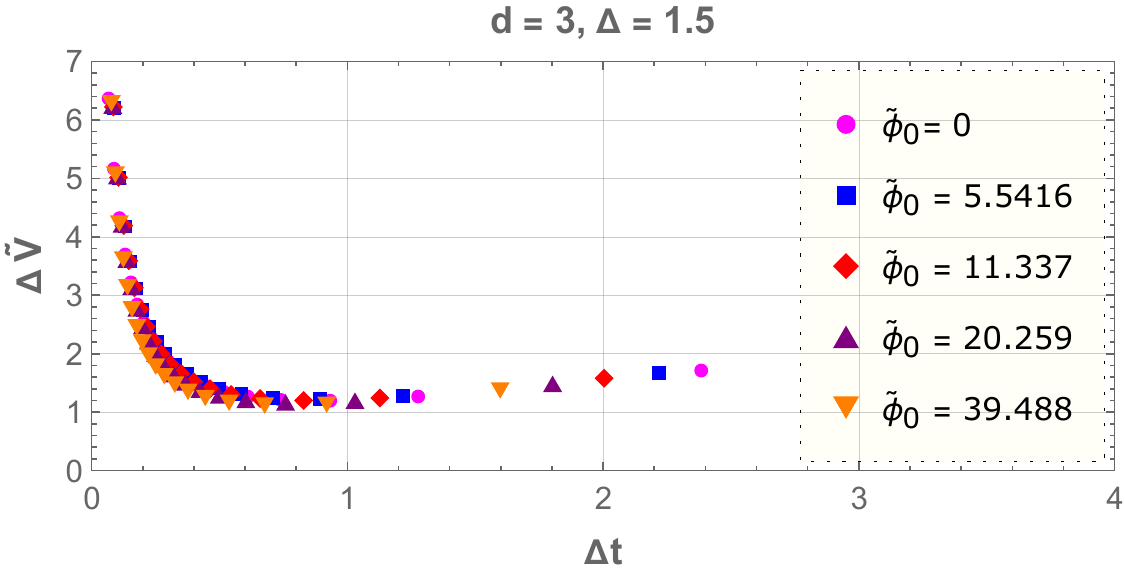}\\
    \caption{$\Delta V$ versus the boundary time interval $\Delta t$ for various boundary deformation parameter $\tilde{\phi}_0=T^{-d+\Delta}\phi_0$.}
    \label{fig:PlotV}
\end{figure*}
In this section, we calculate timelike holographic complexity recently proposed in \cite{Alishahiha2025} which extends the previous holographic subregion complexity proposal \cite{Alishahiha2015} from spacelike to timelike subregions. For a timelike subregion $\mathcal{T}$ in the boundary, the timelike holographic complexity is defined as
\begin{equation}
\mathcal{C}_\mathcal{T}=\frac{V_\mathcal{T}}{G_N^{(d+1)}L},
\end{equation}
where $G_N^{(d+1)}$ is the $(d+1)$-dimensional gravitational constant, $L$ is the AdS radius, and $V_\mathcal{T}$ is a volume enclosed by extremal surfaces corresponding to timelike subregion in the boundary.\\
\indent In \cite{Alishahiha2025}, the timelike holographic complexity is obtained from the volume difference between spacelike and timelike branches. Here, we compute the volume difference between the regions enclosed by the $t_+(r)$ and $t_-(r)$ surfaces, both outside and inside the horizon. The total volume can be calculated from
\begin{align}\label{V_T}
    V_\mathcal{T}&=2V_{d-2}L^{d}\int\frac{t(r)}{r^{d}}dr\nonumber\\
    &=V_{d-2}L^{d}\frac{\Delta t}{({d-1})\varepsilon^{d-1}}+\frac{2V_{d-2}L^{d}}{(d-1)}\int\frac{t'(r)}{r^{d-1}}dr,
\end{align}
where the factor 2 arises from the symmetric nature of the extremal surfaces. We use integration by parts to go from the first line to the second line using $t(0)=\frac{\Delta t}{2}$. Note that the first term of Eq. \eqref{V_T} corresponds exactly to the volume calculated in the pure AdS$_3$ background. \\
\indent In \cite{Alishahiha2025}, it is initially argued that the calculations of an extremal surface correspond to the timelike boundary subregion involves a turning point $r_0$ inside the horizon, or $r_0>r_H$. However, in this work, we consider $r_0<r_H$ as the turning point of the timelike surface that is merged with spacelike surfaces in the interior. Therefore, in this setting, a volume enclosed by an extremal surface homologous to $\mathcal{T}$ is given by the difference between the volume enclosed by $t_+(r)$ and $t_-(r)$ surfaces, which is depicted in figure \ref{fig:CV}. Hence, the corresponding volume subtracted by the leading divergence term is given by
\begin{align}\label{DeltaV}
    \Delta V\equiv &V_\mathcal{T}-\frac{V_{d-2}L^d\Delta t}{(d-1)\varepsilon^{d-1}}\\\nonumber
    =&\frac{2V_{d-2}L^d}{d-1}\bigg[\int_\varepsilon^\infty \frac{t_{+}^\prime(r)}{r^{d-1}}dr-\int_{r_0}^\infty \frac{t_-^\prime(r)}{r^{d-1}}dr\bigg].
\end{align}
It is interesting to see that $\Delta V$ is strictly real-valued. The finite volume $\Delta V$ can be interpreted as the complexity of formation as studied in~\cite{Chapman2017}, namely the additional complexity (or volume, in this framework) required to prepare the thermal state compared to preparing the vacuum state.
\\
\indent As briefly explained in the previous section, we should pick a negative branch of $t^\prime_s(r)$ in the interior to avoid possible extremal surfaces crossing. Therefore, in calculating $\Delta V$, we should separate the $0<r<r_H$ and $r>r_H$ integrals. This will give us
\begin{align}
    \Delta V=&\frac{2V_{d-2}L^{d}}{d-1}\bigg[\int_\varepsilon^{r_H-\delta}\frac{t_+^\prime(r)}{r^{d-1}}-\int_{r_H+\delta }^\infty \frac{t_+^\prime (r)}{r^{d-1}}\\\nonumber
    &-\bigg(\int_{r_0}^{r_{H}-\delta }\frac{t_-^\prime(r)}{r^{d-1}}-\int_{r_H+\delta}^\infty \frac{t_-^\prime(r)}{r^{d-1}}\bigg)\bigg].
\end{align}
All of the integrals exhibit divergences at the horizon and therefore require the introduction of a near–horizon regulator $\delta$. However, these regulator–dependent terms cancel exactly, leaving the total quantity $\Delta V$ divergence-free. Moreover, we explicitly show that, in $d=2$, the resulting expression remains finite and strictly positive in the limit $\varepsilon\rightarrow0$, showing that $\Delta V$ remains free of near-horizon and UV divergences.\\
\indent For BTZ black hole solution with $d=2$ and $\phi(r)=0$, the analytical solution to $\Delta V$ can be expressed as (See Appendix B for the derivation)
\begin{align}
    \Delta V\big|_{\phi_0=0}=- 2L^2\sinh^{-1}&\left(\text{csch}\left(\frac{\Delta t}{2r_H}\right)\right)\nonumber\\ &+2L^2\sinh^{-1}\left(\text{coth}\left(\frac{\Delta t}{2r_H}\right)\right).
\end{align}
The first term is the contribution from the spacelike surface $t_+^\prime(r)$ while the second term comes from the timelike surface $t_-^\prime(r)$. In this case, $\Delta V\rightarrow0$ as $\Delta t\rightarrow0$, grows linearly in $\Delta t$, and saturates to a constant as $\Delta t\rightarrow\infty$. It is worth noting that the growth of $\Delta V$ at late times is directly associated with the choice of the negative branch of $t_s^\prime(r)$ discussed in the previous section. The most significant aspect of this result is that, in the 
$d = 2$ BTZ background, the quantity $\Delta V$ is highly sensitive to the interior solution since all contributions 
associated with the UV cutoff as well as the turning point vanish as $ \varepsilon \to 0 $ (See Appendix B for the derivation), leaving the terms evaluated at $r\rightarrow\infty$ end of the integration.\\
\indent Our BTZ calculations also show that $\mathcal{C}_\mathcal{T}$ does not simply follow from the analytical continuation of its spacelike counterpart derived in \cite{Auzzi2020}. A naive analytic continuation $l\rightarrow i\Delta t$, where $l$ is the length of the spacelike subsystem, would yield a complex-valued $\mathcal{C}_\mathcal{T}$, whereas from the CV prescription, we find that the holographic complexity remains strictly real. Moreover, the complexity for a timelike subregion does not contain a logarithmic divergent term proportional to the entanglement entropy; rather we find a UV-finite result controlled by the interior structure. \\
\indent When the scalar field is turned on, the function $t^\prime_s(r)$ is given by Eq. \eqref{tprimes}, and the integral can be performed numerically using the numerical solutions to $f(r)$ and $\chi(r)$ both in the exterior and interior of the black hole. We calculate the integral of the spacelike surfaces and subtract the result with the integral of the timelike surfaces (see figure \ref{fig:CV}). We then plot the value of 
\begin{equation}
    \Delta\tilde{V}=\frac{(d-1)\Delta V}{V_{d-2}L^d},
\end{equation}
versus $\Delta t$. Here, we choose the same variations of $d,\Delta$, and $r_0$ as the ones used to calculate HTEE. The result can be seen in figure \ref{fig:PlotV}.\\
\indent Again, our numerical calculations reproduce the analytical result for the $d=2$ BTZ case. The scalar field generically increase $\Delta\tilde{V}$ for $d=2$ while it initially decrease the complexity in $d=3$ before eventually coincide and undergo a linear growth in late $\Delta t$. In $d=2$, the linear growth is governed by the interior area contribution while the effect of $\phi_0$ mainly modifies the exterior region.
\begin{figure}
    \centering
    \includegraphics[width=0.47\linewidth]{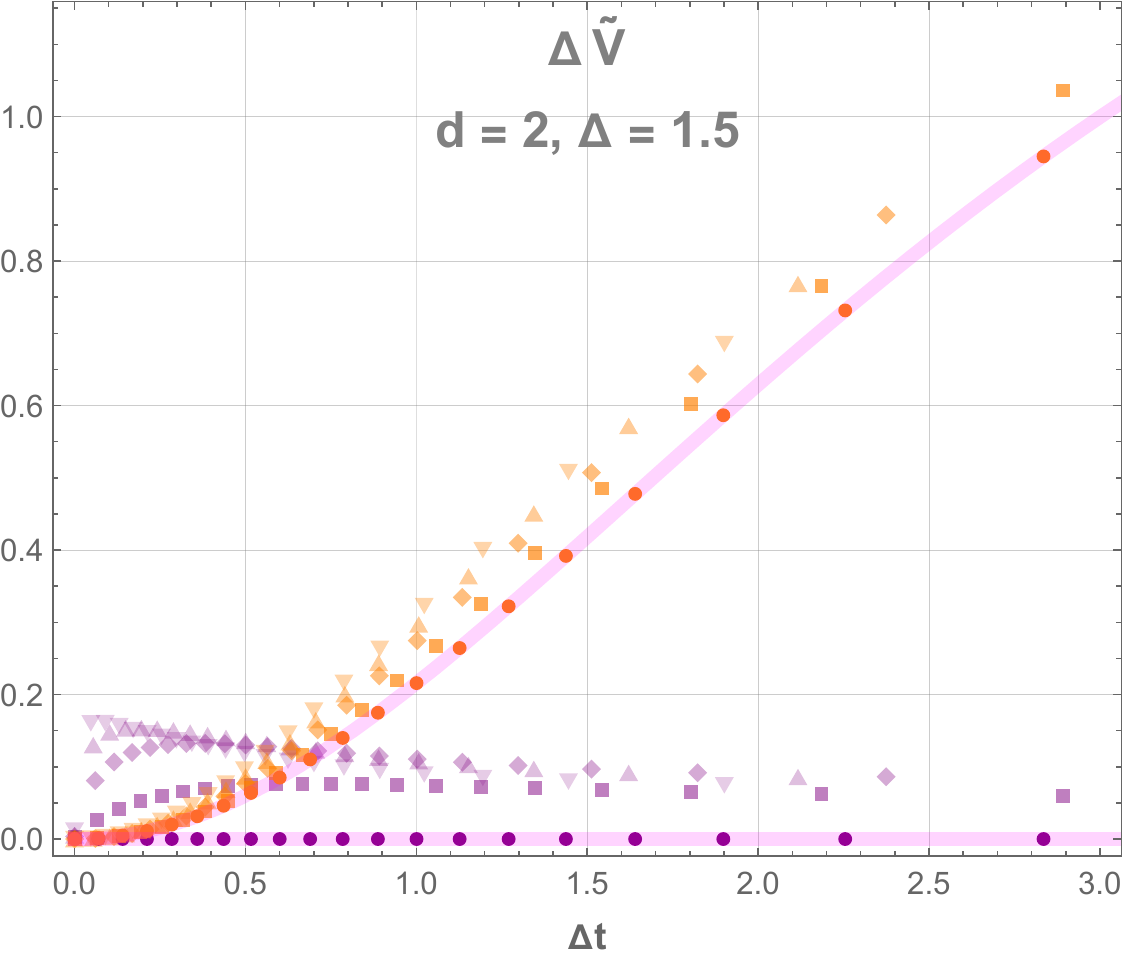}\;
    \includegraphics[width=0.47\linewidth]{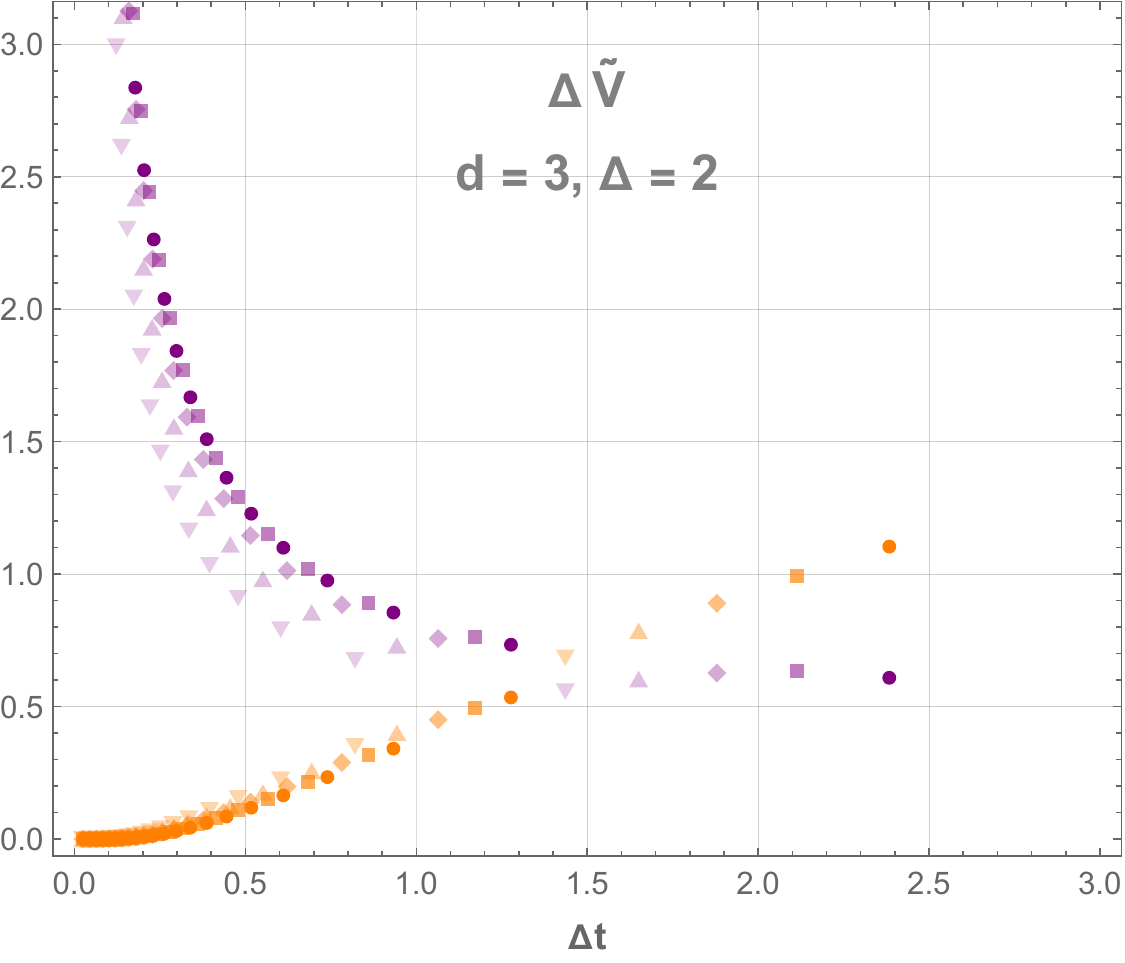}
    \caption{Exterior (purple) vs. interior (orange) contributions to $\Delta\tilde{V}$. Solid magenta lines represent the known analytical solution. Decreasing opacity depicts increasing $\tilde{\phi_0}$.}
    \label{fig:placeholder}
\end{figure}
\section{Discussions}\label{sec5}
In this work, we investigate the HTEE and the holographic timelike subregion complexity in a black hole background with a scalar hair. A key distinction between holographic timelike entanglement entropy and temporal entanglement or standard spacelike entanglement entropy lies in their sensitivity to the bulk geometry and the black hole interior. While temporal and spacelike entanglement entropy in small subregion length limit is often governed by near-boundary data, the HTEE always involves extremal surfaces that extend deep into the black hole interior, even in small $\Delta t$ limit. As a result, the HTEE is sensitive not only to near-boundary data but also to the full bulk solution, including the trans-IR region governed by the interior geometry. This backreacted bulk sensitivity explains why near-boundary expansions alone are insufficient to fully reproduce the HTEE through analytic continuation, although we have shown partial agreement. In this sense, HTEE provides a sharper probe of bulk geometry than its spacelike counterpart.\\
\indent A key outcome of our analysis is the explicit breakdown of the analytic continuation relating spacelike and temporal entanglement entropy to the HTEE in the presence of relevant boundary deformation $\phi_0$. While these two probes are analytically connected in the undeformed theory, we find that the boundary relevant deformation destroys this equivalence, leading to a qualitative mismatch even in $d=2$ or AdS$_3$. This result highlights a qualitative difference between spacelike and timelike entanglement probes in the presence of relevant deformations, indicating that HTEE encodes information beyond what is accessible from spacelike observables. Our results suggest that, in the deformed theory, timelike observables encode information that cannot be reconstructed from analytic continuation of spacelike entanglement alone, highlighting their distinct role in holographic reconstruction. \\
\indent The failure of analytic continuation formulated analytically in small $\Delta t$ limit is an intrinsic consequence of the deformation itself and does not rely on the presence of a black hole, as in this limit, we ignore the presence of the horizon. The influence of the black hole horizon is unavoidable and becomes significant once $\Delta t$ becomes large, and the turning point probe deeper into the bulk. This regime is explored numerically in our analysis.\\
\indent Furthermore, our results show that HTEE from the spacelike-timelike surfaces merger responds in a nontrivial way under the boundary relevant deformation sourced by $\phi_0$, in contrast with the irrelevant deformation induced by the $T\bar{T}$-deformation \cite{JiangWangWuYang2023}, which does not give any effect in finite temperature case. The effect of the $T\bar{T}$ deformation on the bulk geometry can be captured by the introduction of a finite radial cutoff $r_c$ which only modifies the spatial component of the metric. In contrast, the effect of the relevant deformation $\phi_0$ with $\Delta <d$ influences both time and spatial directions through the functions $\chi(r)$ and $f(r)$. From the boundary perspective, this difference arises from the fact that $T\bar{T}$ is an irrelevant deformation that primarily alters the UV regime, whereas the scalar deformation with $\Delta <d$ is relevant and reshapes the IR physics, creating an increasingly large backreaction from the boundary up to the Kasner singularity.\\
\indent An explicit field-theoretic understanding of timelike entanglement entropy under boundary relevant deformations remains an important open problem, since our analysis is performed in the holographic perspective. A direct CFT calculation of the timelike entanglement entropy under a relevant scalar deformation, especially in $d=2$, would provide a valuable comparison to the conjectured holographic calculations. The dependence of timelike entanglement entropy on the boundary time interval $\Delta t$ in the presence of a relevant deformation was anticipated in \cite{XuGuo2025}. It was argued that, for a generic state, the commutation relation between the twist operator and its time derivative can induce an explicit $\Delta t$ dependence, even in $d=2$. In this work, we provide a holographic realization of this effect by explicitly demonstrating the $\Delta t$ dependence of the holographic timelike entanglement entropy under boundary deformation in $d=2$.\\
\indent We also briefly discuss HTEE in black holes with inner horizon, such as the Reissner-Nordstr\"om-AdS black hole. In this setting, the spacelike extremal surface is glued at both outer and inner horizon, as does the timelike counterpart. Those spacelike and timelike surfaces are then merged inside the inner horizon. We have also illustrated qualitatively how the extremal surface behaves as the interior undergoes an ER bridge collapse (destroying the inner horizon) once the scalar deformation is turned on. Although the spacelike-timelike merger could still occur inside the inner horizon due to the universal Kasner-like interior behavior, an understanding of how the ER bridge collapse and the possibility of the Kasner inversion/transition affect the structure of extremal surfaces is left for future works.\\
\indent From the discussions of the extremal surfaces inside multi-horizon black holes, it is noted that a different branch for $t_s^\prime(r)$ is required in between the outer and the inner horizon to avoid crossing between surfaces. Certainly, this will not affect the area of the extremal surface, since it only depends on $(t^\prime_s)^2$. However, it is very interesting to see that this significantly affects the volume enclosed by the spacelike and timelike extremal surfaces, in order to calculate holographic timelike subregion complexity in the CV perspective. The choice of the non-crossing branch leads to an increasing complexity in time observed in $d=2$, consistent with the expected behavior of quantum computational complexity that increases over time. This shows that holographic complexity could probe aspects of the interior geometry that remain invisible to entanglement entropy. \\
\indent Applying this analysis to the BTZ case, we find that the subleading contribution to the holographic complexity exhibits a linear growth in $\Delta t$ at early times before eventually saturating. We also show that the linear growth in $\Delta t$ is primarily governed by the volume inside the horizon. Due to this choice as well, we find that the scalar field generically increases the subregion complexity, which is the expected result from the boundary point of view. Naively, relevant deformations on the boundary would break the conformal symmetry, and this would imply that the final quantum state is harder to prepare.
\section*{Acknowledgement}
We would like to thank Mir Afrasiar and Jaydeep Kumar Basak for useful discussions. HLP is financially supported by National Research and Innovation Agency (BRIN) through its postdoctoral program. FK would like to thank the Ministry of Primary and Secondary Education (Kemendikdasmen) for the financial support through the Beasiswa Unggulan program. FPZ would like to thank PPMI FMIPA ITB for financial support. HLP, FK, and FPZ would like to thank members of theoretical physics group, Institut Teknologi Bandung, for useful discussions.
\section*{Appendix A: Extremal Surfaces for BTZ Black Holes}
In this appendix, we rederive the expression of the extremal surfaces for the three-dimensional BTZ black hole which given by the equation ~\eqref{analyticalarea} using the method developed in \cite{Afrasiar2025,afrasiar2025BH}. Before deriving the expression of the extremal surfaces, we first determine the expression for $t(r)$. By substituting $f(r)=1-\big(\frac{r}{r_H}\big)^2$ and $\chi(r)=0$ into Eq.~\eqref{tprimesquare}, we obtain
\begin{align}
       t^{\prime 2}(r)&=\frac{K^2 r^{2}}{f^2}\frac{1}{f+K^2r^{2}}\nonumber\\
    &=\frac{K^2 r^{2}}{(r_H^2-r^2)^2}\frac{r_H^{6}}{r_H^2-r^2+K^2r_H^{2}r^{2}}. 
\end{align}
As explained in \cite{Afrasiar2025} (or the paragraph above Eq~\eqref{Ktilde}), we onsider for the case where $K^2=\pm\tilde{K}^2$, where ${K}^2$ is a positive real number. Consider the case $K^2=\tilde{K}^2$ first. In this case, $t^{\prime}(r)=t_+^{\prime}(r)$ will be given by
\begin{equation}
     t_+^{\prime }(r)=\frac{\tilde{K} r}{(r_H^2-r^2)}\frac{r_H^{3}}{\sqrt{r_H^2+(\tilde{K}^{2}r_H^2-1)r^{2}}}.
\end{equation}
Then the function $t_+(r)$ will be given by the integral
\begin{align}
    t_+(r)=\int dr \frac{\tilde{K}r}{(r_H^2-r^2)}\frac{r_H^3}{\sqrt{r_H^2+(\tilde{K}^{2}r_H^2-1)r^{2}}}.
\end{align}
To evaluate the integral, we do substitution $u^2=r_H^2+(\tilde{K}^{2}r_H^2-1)r^{2}$. Then, the integral will simplify into
\begin{equation}
    t_+(r)=\int du \frac{\tilde{K}r_H^3}{\tilde{K}^2r_H^4-u^2},
\end{equation}
which can be solved by doing another substitution. The solution of this integral is
\begin{equation}
     t_+(r)=\frac{r_H}{2}\ln\left[\frac{\tilde{K}r_H^2+\sqrt{r_H^2+(\tilde{K}^2r_H^2-1)r^2}}{\tilde{K}r_H^2-\sqrt{r_H^2+(\tilde{K}^2r_H^2-1)r^2}}\right] + C_1,
\end{equation}
where $C_1$ is a constant of integration that can be determined by the initial condition where $t_+(r=0)=\frac{\Delta t}{2}$. Using this condition, we obtain that
\begin{equation}
    C_1=\frac{\Delta t}{2}-\frac{r_H}{2}\ln\left[\frac{\tilde{K}r_H+1}{\tilde{K}r_H-1}\right]
\end{equation}

Next, consider the case where $K^2=-\tilde{K}^2$. In this case, $t^{\prime}(r)=t_-^{\prime}(r)$ will be given by
\begin{equation}
     t_-^{\prime 2}(r)=\frac{\tilde{K}^2 r^{2}}{(r_H^2-r^2)}\frac{r_H^{3}}{(\tilde{K}^{2}r_H^2+1)r^{2}-r_H^2},
\end{equation}
and $t_-(r)$ will be given by the integral
\begin{equation}
     t_-^{\prime }(r)=\frac{\tilde{K} r}{(r_H^2-r^2)}\frac{r_H^{3}}{\sqrt{(\tilde{K}^{2}r_H^2+1)r^{2}-r_H^2}}.
\end{equation}
In order to have a real value for $t_-(r)$, $r$ must satisfy the condition $r>\frac{r_H}{\sqrt{1+\tilde{K}^2r_H^2}}$. This condition gives us the definition of turning point $r_0$, where near this point $t_-^{\prime}(r_0)\to\infty$. The expression for $t_-(r)$ can be obtained in the same manner as the one for $t_+(r)$.Thus, we obtain
\begin{align}
    t_-(r)&=\int dr \frac{\tilde{K}r}{(r_H^2-r^2)}\frac{r_H^3}{\sqrt{(\tilde{K}^{2}r_H^2+1)r^{2}-r_H^2}}\nonumber\\
    &=\frac{r_H}{2}\ln\left[\frac{\tilde{K}r_H^2+\sqrt{r_H^2+(\tilde{K}^2r_H^2+1)r^2}}{\tilde{K}r_H^2-\sqrt{r_H^2+(\tilde{K}^2r_H^2+1)r^2}}\right] + C_2,
\end{align}
where $C_2$ is an another constant of integration, fixed by the condition $t_-(r_0)=0$, from which we obtain $C_2=0$. 
The constant $\tilde{K}$ can be related to time interval $\Delta t$ by identifying $t_+(r)$ with $t_-(r)$ at the infinity by imposing the condition $t_+(\infty)=t_-(\infty)$. With this condition, the relation between $\tilde{K}$ and $\Delta t$ is given by
\begin{equation}\label{Ktrelation}
    \tilde{K}=\frac{1}{r_H}\coth{\left(\frac{\Delta t}{2r_H}\right)}.
\end{equation}
Identifying $\tilde{K}^2$ as $\frac{f(r_0)}{r_0^2}$ where $r_0$ is the turning point of the imaginary surface, we have the relation
\begin{equation}
    \Delta t=2r_H\tanh^{-1}\bigg(\frac{(r_0/r_H)}{\sqrt{1-(r_0/r_H)^2}}\bigg).
\end{equation}
\indent Now, we will derive the expression of the extremal surfaces for of the three dimension BTZ black. In this case, by substituting equation ~\eqref{tprimesquare} The equation ~\eqref{areafunctional} can be rewritten as
\begin{align}
    \mathcal{A}_s(\gamma_\mathcal{T})&=2L\int\frac{dr}{r}\frac{1}{\sqrt{f+s\tilde{K}^2r^2}}\nonumber\\
    &=2L\int\frac{dr}{r}\frac{r_H}{\sqrt{r_H^2-r^2+s\tilde{K}^2r_H^2r^2}},
\end{align}
where $s=\pm1$. For $s=+1$, The extremal surfaces is given by the integral
\begin{align}
     \mathcal{A}_+(\gamma_\mathcal{T})=2L\int_{\epsilon}^{\infty}\frac{dr}{r}\frac{r_H}{\sqrt{r_H^2+(\tilde{K}^2r_H^2-1)r^2}}.
\end{align}
This integral can be evaluated using the same substitution for calculating $t_+(r)$, which simplifies the integral into
\begin{equation}
    \mathcal{A}_+(\gamma_\mathcal{T})=2L\int_{u(\epsilon)}^{\infty}du\frac{r_H}{u^2-r_H^2},
\end{equation}
and by using another substitution gives us
\begin{align}
    \mathcal{A}_+(\gamma_\mathcal{T})&=-L\ln{\left[\frac{\sqrt{r_H^2+(\tilde{K}^2r_H^2-1)\epsilon^2}-r_H}{\sqrt{r_H^2+(\tilde{K}^2r_H^2-1)\epsilon^2}+r_H}\right]}\nonumber\\
    &\simeq 2L\ln{\left(\frac{2r_H}{\epsilon}\frac{1}{\sqrt{\tilde{K}^2r_H^2-1}}\right)}\nonumber\\
    &=2L\ln{\left(\frac{2r_H}{\epsilon}\sinh\left(\frac{\Delta t}{2r_H}\right)\right)},
\end{align}
where we have used Eq~\eqref{Ktrelation} when going to the third line. For the case $s=-1$, the extremal surfaces takes the form
\begin{equation}
     \mathcal{A}_-(\gamma_\mathcal{T})=-2iL\int_{r_0}^{\infty}\frac{dr}{r}\frac{r_H}{\sqrt{(\tilde{K}^2r_H^2+1)r^2-r_H^2}},
\end{equation}
which can be simplified and evaluated by doing the same substitution as in $t_-(r)$. After performing the substitution, we obtain
\begin{align}
    \mathcal{A}_-(\gamma_\mathcal{T})=-2iL\int_{0}^{\infty}du\frac{r_H}{u^2+r_H^2}=i\pi L.
\end{align}
We may extend the discussion by further asking what happens when one tries to compute the area separately in two different regions, namely, the exterior and the interior part. In fact, it can be shown that the interior part of $\mathcal{A}_+(\gamma_\mathcal{T})$ is always linearly proportional to $\Delta t$.
\begin{align}
     \mathcal{A}_{\text{int},+}(\gamma_\mathcal{T})&=2L\int_{r_H+\epsilon}^{\infty}\frac{dr}{r}\frac{r_H}{\sqrt{r_H^2+(\tilde{K}^2r_H^2-1)r^2}} \nonumber \\ &= -L\ln{\left[\frac{\sqrt{r_H^2+(\tilde{K}^2r_H^2-1)(r_H+\epsilon)^2}-r_H}{\sqrt{r_H^2+(\tilde{K}^2r_H^2-1)(r_H+\epsilon)^2}+r_H}\right]}\nonumber \\  &= L\ln{\left[\frac{\tilde{K}r_H^2+r_H}{\tilde{K}r_H^2-r_H}\right]}\nonumber \\  &= \frac{L\Delta t}{r_H}.
\end{align}

\section*{Appendix B: Subregion Complexity for BTZ Black Holes}
\begin{figure}[h]
    \centering
    \includegraphics[width=0.7\linewidth]{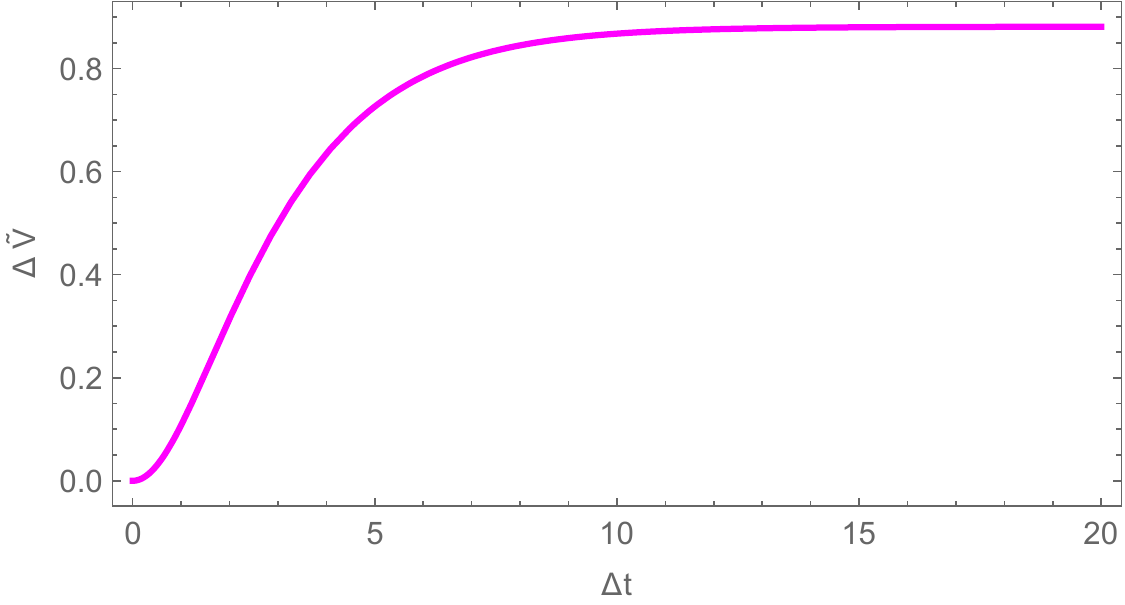}
    \caption{$\Delta\tilde{V}$ vs. $\Delta t$ in BTZ background.}
    \label{fig:BTZComplexity}
\end{figure}

This appendix considers the analytical expression of $\Delta V$ in Eq.~\eqref{DeltaV} under the BTZ black hole background ($d=2$). By employing Eq.~\eqref{tprimes}, the first integral can be computed as follows
\begin{align}
    \int_\epsilon^\infty\frac{t_+'(r)}{r}dr&=\int_\epsilon^\infty \frac{dr}{rf}\left(1+\left(\frac{r_0}{r}\right)^2\frac{f}{f_0}\right)^{-1/2}\nonumber \\ &=-\frac{1}{2}\int \frac{du}{u\sqrt{1+(r_0^2/f_0)u}}
\end{align}
where $u$ is defined as $u:=f/r^2$. The integral can be performed, and we would get the following expression
\begin{align}
    \int_\epsilon^\infty\frac{t_+'(r)}{r}dr&= \frac{1}{2}\ln\left|\frac{\sqrt{1+(r_0^2/f_0)u}+1}{\sqrt{1+(r_0^2/f_0)u}-1}\right|_{u=\frac{1}{\epsilon^2}-\frac{1}{r_H^2}}^{u=-\frac{1}{r_H^2}} \nonumber \\ &=\tanh^{-1}\left(\sqrt{1-\frac{r_0^2}{f_0r_H^2}}\right) \nonumber \\ &=\tanh^{-1}\left(\sqrt{\frac{1-2y^2}{1-y^2}}\right)
\end{align}
where $y$ is defined as the ratio of $r$ when it is evaluated at the turning point to when it is evaluated at the horizon, $y := r_0 / r_H$. Likewise, for the timelike surface, the integral can be computed as follows
\begin{align}
    \int_{r_0}^\infty\frac{t_-'(r)}{r}dr&=\int_{r_0}^\infty \frac{dr}{rf}\left(1-\left(\frac{r_0}{r}\right)^2\frac{f}{f_0}\right)^{-1/2}\nonumber \\ &=-\frac{1}{2}\int \frac{du}{u\sqrt{1-(r_0^2/f_0)u}} \nonumber \\&= \frac{1}{2}\ln\left|\frac{\sqrt{1-(r_0^2/f_0)u}+1}{\sqrt{1-(r_0^2/f_0)u}-1}\right|_{u=\frac{1}{{r_0}^2}-\frac{1}{r_H^2}}^{u=-\frac{1}{r_H^2}} \nonumber \\ &=\coth^{-1}\left(\frac{1}{\sqrt{1-y^2}}\right)
\end{align}
It needs to be noted that for both integrals, the terms coming from the UV cutoff and the turning point vanish automatically after taking the $\epsilon\to 0$ limit. Meaning that the subleading term of the total volume is sensitive to the interior part of the BTZ black hole geometry. We can rewrite $\Delta V$ as a function of the boundary time $\Delta t$ by using Eq.~\eqref{dtanalytical}. After performing the algebra, we obtain the following expression
\begin{align}\label{DeltaVAppendix}
    \Delta V= 2L^2\sinh^{-1}&\left(\text{coth}\left(\frac{\Delta t}{2r_H}\right)\right)\nonumber\\ &-2L^2\sinh^{-1}\left(\text{csch}\left(\frac{\Delta t}{2r_H}\right)\right).
\end{align}
The plot of $\Delta \tilde{V}=\frac{\Delta V}{L^2}$ versus $\Delta t$ for BTZ background can be seen in figure \ref{fig:BTZComplexity}. It is interesting to note that $\Delta \tilde{V}$ starts to saturate at thermal time, namely when $\Delta t \sim \mathcal{O}(r_H)$. The saturation at late times $\Delta t \to \infty$ is consistent with the results in~\cite{Chapman2017}, where it was shown that for $d=2$, the complexity of formation is found to be constant, in contrast to higher dimensions. The main difference between this prescription and~\cite{Chapman2017} is that we are able to determine the dependence of $\Delta \tilde{V}$ on the time $\Delta t$.
\\
\indent It also needs to be noted that after taking the small $\Delta t$ approximation of Eq.~\eqref{DeltaVAppendix}, one obtains a quadratic growth, i.e., $\Delta V\sim (\Delta t)^2$. Following the initial quadratic growth, $\Delta V$ enters a linear-growth regime. After a time on the order of the thermal timescale, the finite volume finally saturates.
\indent The previous discussion may not be fully satisfactory since we have not discussed the possible divergence at the horizon. To analyze the term at the horizon, we may separate the integral into the exterior part and the interior part. Let us calculate the exterior part of $\Delta V$.
\begin{align}
    \int_\epsilon^{r_H-\epsilon}\frac{t_+'(r)}{r}dr&= \frac{1}{2}\ln\left|\frac{\sqrt{1+(r_0^2/f_0)u}+1}{\sqrt{1+(r_0^2/f_0)u}-1}\right|_{u=\frac{1}{\epsilon^2}-\frac{1}{r_H^2}}^{u=\frac{2\epsilon}{r_H^3}} \nonumber \\ &=\frac{1}{2}\ln\left|\frac{\sqrt{1+(r_0^2/f_0)(2\epsilon/r_H^3)}+1}{\sqrt{1+(r_0^2/f_0)(2\epsilon/r_H^3)}-1}\right|,
\end{align}
and we also have
\begin{align}
    \int_\epsilon^{r_H-\epsilon}\frac{t_-'(r)}{r}dr&= \frac{1}{2}\ln\left|\frac{\sqrt{1+(r_0^2/f_0)u}+1}{\sqrt{1+(r_0^2/f_0)u}-1}\right|_{u=\frac{1}{\epsilon^2}-\frac{1}{r_H^2}}^{u=\frac{2\epsilon}{r_H^3}} \nonumber \\ &=\frac{1}{2}\ln\left|\frac{1+\sqrt{1-(r_0^2/f_0)(2\epsilon/r_H^3)}}{1-\sqrt{1-(r_0^2/f_0)(2\epsilon/r_H^3)}}\right|.
\end{align}
Hence, we obtain
\begin{align}
    \Delta V_{\text{ext}}&=\frac{1}{2}\ln\left|\frac{1+\sqrt{1-(r_0^2/f_0)(2\epsilon/r_H^3)}}{1-\sqrt{1-(r_0^2/f_0)(2\epsilon/r_H^3)}}\right|\nonumber \\ &\quad-\frac{1}{2}\ln\left|\frac{\sqrt{1+(r_0^2/f_0)(2\epsilon/r_H^3)}+1}{\sqrt{1+(r_0^2/f_0)(2\epsilon/r_H^3)}-1}\right| \nonumber \\ &=0.
\end{align}
The result is equal to zero in the limit \(\epsilon \to 0\). As a consequence, the interior part gives the entire contribution to \(\Delta V\). Namely, \(\Delta V_{\text{int}} = \Delta V\) in $d=2$.
\bibliography{bib.bib}
\end{document}